\documentclass[fleqn,usenatbib]{mnras}

\usepackage{newtxtext,newtxmath}
\usepackage[T1]{fontenc}
 \usepackage{comment}
\DeclareRobustCommand{\VAN}[3]{#2}
\let\VANthebibliography\thebibliography
\def\thebibliography{\DeclareRobustCommand{\VAN}[3]{##3}\VANthebibliography}

\newcommand{\msun}{\,\rm M_\odot}
\definecolor{ejcol}{rgb}{0,0,0}
\def\ej#1{\textcolor{ejcol}{{#1}}}

\defcitealias{Shin+2025}{Paper~I}

\usepackage{graphicx}	% Including figure files
\usepackage{amsmath}	% Advanced maths commands
\usepackage{pifont}% http://ctan.org/pkg/pifont
\usepackage{hyperref}
\usepackage{dblfloatfix}
\usepackage{placeins}

\title[Impact of stellar feedback on IMBH growth]{The MandelZoom project II: the impact of stellar feedback on black hole accretion through an $\alpha$-disc in dwarf galaxies with a resolved interstellar medium }

\author[E. -J. Shin et al.]{
Eun-jin Shin,$^{1,2}$\thanks{E-mail: ejs237@cam.ac.uk}
Matthew C. Smith,$^{3}$
Debora Sijacki,$^{1,2}$
Martin A. Bourne,$^{2,4}$ and
Sophie Koudmani $^{4,5}$
\\
$^{1}$Institute of Astronomy, University of Cambridge, Madingley Road, Cambridge CB3 0HA, UK\\
$^{2}$Kavli Institute for Cosmology, University of Cambridge, Madingley Road, Cambridge CB3 0HA, UK\\
$^{3}$Max-Planck-Institut f{\"u}r Astrophysik, Karl-Schwarzschild-Str. 1, D-85748, Garching, Germany\\
$^{4}$Centre for Astrophysics Research, Department of Physics, Astronomy and Mathematics, University of Hertfordshire, College Lane, Hatfield, AL10 9AB, UK\\
$^{5}$St Catharine's College, University of Cambridge, Trumpington Street, Cambridge CB2 1RL, UK\\
}

\date{MNRAS, submitted}

\pubyear{\the\year{}}

\begin{document}
\label{firstpage}
\pagerange{\pageref{firstpage}--\pageref{lastpage}}
\maketitle
%246
\begin{abstract}
We present a suite of high-resolution simulations to study how different stellar feedback channels regulate the growth of central intermediate-mass black holes (IMBHs) in dwarf galaxies hosting nuclear star clusters (NSCs). We employ a super-Lagrangian refinement scheme to resolve the self-gravity radius of the $\alpha$-accretion disc ($<0.01$~pc) and follow the gas inflows from the interstellar medium (ISM) to the black hole (BH), allowing for the self-consistent emergence of circumnuclear discs (CNDs). In the absence of stellar feedback, as expected, the galactic disc fragments excessively, producing a massive CND. When radiative stellar feedback is included, fragmentation is suppressed, with even more massive CNDs forming and feeding the IMBH. With supernova (SN) feedback only, clustered SNe strongly heat the ISM, yielding both the lowest CND masses and BH accretion rates. When both radiative stellar feedback and SNe are included, the CND becomes intermittent: it survives for $10$--$100$~Myr, and is then destroyed by feedback before being replenished by fresh galactic inflows, while substantial BH growth still takes place. These results highlight the critical importance of accurately modelling the combined effects of key stellar feedback processes to understand IMBH growth. Our simulation suite brackets the likely range of CND states, with IMBHs exhibiting significant growth and systematic spin-up in all dwarf galaxy models explored. These findings bode well for the detection of IMBHs with future observational facilities such as SKA, the Rubin Observatory, and LISA, and make them highly relevant progenitor candidates of the high-redshift supermassive BHs observed by JWST.
\end{abstract}

% Select between one and six entries from the list of approved keywords.
% Don't make up new ones.

%%%%%%%%%%%%%%%%%%%%%%%%%%%%%%%%%%%%%%%%%%%%%%%%%%

%%%%%%%%%%%%%%%%% BODY OF PAPER %%%%%%%%%%%%%%%%%%

\section{Introduction}

Unveiling the origin of supermassive black holes (SMBHs) and their coevolution with galaxies remains a fundamental question in astrophysics \citep[see e.g., seminal papers by][]{Rees1984, Efstathiou+Rees1988,  Kormendy+Richstone1995, Silk+Rees1998, King2003}. Intermediate-mass black holes (IMBHs; $10^2$--$10^{5-6}\,\msun$) may hold a key to understanding this process, unravelling if the coevolution holds all the way down to the dwarf galaxy regime, as well as providing us a unique insight into how this coevolutionary picture may have emerged at early cosmic times. A central challenge here is to determine how IMBHs may form, how abundant they are, and under which physical conditions they can efficiently accrete gas. Furthermore, clarifying when and how such accretion episodes make IMBHs observable, for example, through active galactic nucleus (AGN) activity, mergers with other compact objects, tidal disruption events (TDEs), or feedback, is essential for constraining black hole (BH)–galaxy coevolution \citep[e.g.,][]{Wang+Merritt2004, Mezcua2017, Greene+2020, Koudmani+2022, Pucha+2025}.

Several studies suggest that the growth of light seeds, originating from Pop III remnants \citep[see e.g.,][]{Hirano+2014}, is generally inefficient within the shallow potential wells of their host mini-haloes and may be strongly suppressed by stellar and BH feedback \citep{Johnson+Bromm2007, SmithB+2018, Sassano+2023}, while their small masses make them prone to wandering \citep{Pfister+2019}, potentially stranding them in off-centre orbits that prevent efficient gas accretion \citep[for a review see e.g.,][]{Inayoshi+2020}.

By contrast, more massive stellar systems such as nuclear star clusters (NSCs) offer particularly favourable cradles for IMBH formation and growth. NSCs are ubiquitous even in low-mass galaxies \citep{Sanchez-Janssen+2019, Neumayer+2020, Hoyer+2021, Poulain+2025}, though their formation pathways remain under active investigation \citep[e.g.,][]{Fahrion+2022a, Fahrion+2022b, vanDonkelaar+2024, Lahen+2025, Gray+2025}. Recent $N$-body simulations further demonstrate that IMBHs can form within massive clusters in only a few to several tens of Myr, particularly under low-metallicity conditions \citep[e.g.,][]{Katz+2015, Fujii+2024, Rantala+2024, Rantala+2025}. These results highlight several physical mechanisms that make NSCs uniquely conducive to BH growth: (i) mass segregation and dynamical friction drive hierarchical BH mergers; (ii) their deep potentials, in the case of massive clusters, help BH retention against dynamical ejections from two- and three-body interactions as well as gravitational-wave recoil kicks \citep[e.g.,][]{Rantala+2024,  Rantala+2025}; (iii) their dense stellar backgrounds fuel BHs through TDEs \citep[see e.g.,][]{Lee+2023, Rantala+2024, Rantala+2025, Chang+2025}; (iv) they channel gas inflows \citep{Partmann+2025} and promote circumnuclear disc (CND) formation \citep{Shin+2025}; and (v) the NSC, absorbing angular momentum exchange with the CND, facilitates the circularisation of inflowing gas into an accretion disc and thereby promotes the growth of BH mass and spin \citep{Shin+2025}.

Understanding BH accretion requires an accurate treatment of stellar feedback, since the state of the ISM, which is the reservoir of BH fuel, is highly sensitive to the feedback processes incorporated in simulations. Recently, in the context of dwarf galaxies, more realistic stellar feedback models have been implemented in high-resolution (isolated) galaxy simulations, successfully capturing a multiphase ISM and showing that early feedback processes, including stellar winds, photoionisation, and photoelectric heating, strongly regulate star formation by diffusing the ISM and suppressing SN clustering \citep[e.g.,][]{Kimm+2018, Smith+2021, Hislop+2022, Andersson+2024, Deng+2024}. Building on these models, subsequent studies have begun to explore their impact on BH accretion in isolated galaxy simulations that resolve individual feedback channels, showing that BHs in low-mass galaxies accrete intermittently due to rapidly time-varying stellar feedback \citep[e.g.,][]{Sivasankaran+2022, Partmann+2025, Petersson+2025, Shin+2025}. On smaller, cluster scales, the role of stellar winds from Wolf–Rayet (WR) stars in directly fuelling IMBH accretion has been examined through magnetohydrodynamic simulations, which reveal that turbulence and outflows from high-velocity wind collisions remove most of the gas, resulting in low accretion efficiency \citep{Labaj+2025}.

One of the primary stages where such complex gas–star–BH interactions unfold is the CND. On scales of tens of parsecs, CNDs are found within NSCs and act as reservoirs of gas surrounding the central BH, having long been a major focus of both observational and theoretical studies. The CND around Sgr A* has been extensively investigated over the past decades \citep[][and for the review, \cite{Bryant+Krabbe2021}]{Becklin+1982, Genzel+1985, Guesten+1987, Sutton+1990}, and intense nuclear starbursts \citep{Davies+2007, Hicks+2013} and gravitational instability of CNDs \citep{Izumi+2023} have been observed in nearby Seyfert galaxies. Theoretical and numerical efforts have also sought to elucidate the role of CNDs in the life cycles of galactic nuclei. Hydrodynamic simulations show that CNDs are gravitationally unstable, leading to the formation of clumps and stars, followed by partial dispersal driven by stellar feedback \citep{Roskar+2015, Schartmann+2018, Solanki+2023, Barna+2025, Shin+2025}. Under starburst conditions, SN feedback and turbulence can transform a thin, rotation-supported disc into a turbulent, clumpy, geometrically thick torus, which contributes significantly to nuclear obscuration and facilitates gas inflow toward the SMBH \citep{Wada+Norman2002, Wada+2009, Dinh+2021, Shin+2025}.

Despite significant progress, the formation mechanism and long-term fate of the CND remain highly debated. Some studies argue that the CND is a transient structure, repeatedly disrupted by energetic processes such as SN explosions, powerful stellar winds from OB and WR stars, AGN jets, and turbulence driven by toroidal magnetic fields \citep{Mezger+1989, Morris+Serabyn1996, Genzel+2010, Requena-Torres+2012, Lau+2013, Hsieh+2018, Wada+2023, Solanki+2023}. In contrast, other works suggest that the CND can persist as a quasi-stable structure for up to $\sim$10 Myr \citep{Vollmer+Duschl2001, Christopher+2005, Dinh+2021}. The persistence of the CND is closely tied to large-scale galactic inflows and the interaction between the ISM and the nuclear environment, with observations of inflowing gas streams indicating that external gas supply plays a crucial role in sustaining the disc \citep[e.g.,][]{Vollmer+Duschl2001, Liu+2012, Hsieh+2017}. Understanding this complex interplay between star formation, stellar feedback, and CND evolution is therefore essential for constraining the growth of central BHs, and it is equally important to place these processes within the broader galactic context rather than considering the CND in isolation. However, simulations that simultaneously capture a realistic multiphase ISM, trace the long-term CND cycle, and resolve the self-gravitating accretion disc feeding the SMBH remain exceedingly rare \citep{Hopkins+2024a, Shin+2025}.

In {\sc MandelZoom I} \citep{Shin+2025}, hereafter Paper I, we presented a suite of high-resolution simulations of dwarf galaxies that resolve scales from the galactic environment down to the vicinity of the central BH. With a spatial resolution of $\sim$1.75 pc in the outer galactic regions and super-Lagrangian refinement reaching a spatial resolution of sub-0.01~pc near the BH, the self-gravitating radius of the BH's accretion disc was well resolved, allowing us to track the transfer of mass and angular momentum from the galactic scales all the way to the $\alpha$-disc. 
In this work, we build on \citetalias{Shin+2025} by systematically exploring how different forms of stellar feedback modify the ISM and influence the formation of the CND, as well as affect fragmentation and star formation, and ultimately regulate the mass and spin evolution of the BH.

This work is organised as follows. Section~\ref{sec2:methods} describes the simulation setup. In Section~\ref{sec3:gas-proj}--\ref{sec:3d}, we discuss how the ISM evolves with different stellar (feedback) models and hydro setting. We then investigate how CND evolves in different models in Section~\ref{sec:CNDcycle}. Finally, we explore the BH growth in Section~\ref{sec:BHgrowth}. We discuss our results in Section~\ref{sec4:discussion}, highlighting the limitations of our model and potential future improvements before summarising our findings in Section~\ref{sec5:conclusion}.

\section{Numerical Methods}
\label{sec2:methods}
\subsection{Initial conditions}
\label{sec2:IC}
We largely adopt the initial conditions (ICs), baryonic physics, and resolution scheme from \citetalias{Shin+2025}, but for completeness, we briefly summarise the simulation setup here. 

We initialise an isolated dwarf galaxy with a total mass of $10^{10}~\msun$, following a Wolf–Lundmark–Melotte (WLM)-like system described in \citet{Smith+2021}. The system is generated using {\sc MakeNewDisk} \citep{Springel+2005}, and consists of a dark matter halo and a baryonic disc containing both gas and stars. The dark matter halo is modelled with a \citet{Hernquist1990} density profile, adopting a concentration parameter of $c=15$, a spin parameter of $\lambda=0.035$, and a virial radius of $41$~kpc. The baryonic disc has an exponential surface density profile with a radial scale length of $1.1$~kpc. The stellar component has a Gaussian vertical structure with a scale height of $0.7$~kpc. The stellar disc has a total mass of $9.75\times10^6\msun$. The gaseous disc is initialised with a total mass of $6.83\times10^7\msun$, a uniform metallicity of $0.1Z_\odot$, and a temperature of $10^4$~K.  We set a target gas cell mass to $m_{\rm gas, target}= 20~\msun$, enforcing refinement to keep cell masses within a factor of two of this value (outside of the central refinement region). Old disc star and dark matter particles have fixed masses of $20~\msun$ and $1640~\msun$, respectively. Gravitational softenings are adaptive for gas, with a minimum of $1.75$~pc, and fixed at $1.75$~pc and $20~{\rm pc}$ for disc/newly-formed stars and dark matter particles, respectively. To suppress artificial starbursts during the initial dynamical relaxation, the system is evolved for 100~Myr with radiative cooling and turbulence driving enabled, but with star formation turned off.

Unlike many previous isolated dwarf galaxy simulations, our setup explicitly includes a circumgalactic medium (CGM). The CGM is characterised by radial profiles in temperature, pressure, velocity, and metallicity, informed by high-resolution cosmological zoom-in simulations of dwarf galaxies from \citet{Koudmani+2022}. Within the halo virial radius, the CGM follows the \citet{Hernquist1990} density profile, enclosing a total mass of $2.64\times10^7~\msun$. Beyond this radius, the gas is set to a uniform background with a hydrogen number density of $n_{\rm H} = 2.24\times10^{-6}\,{\rm cm}^{-3}$ and a temperature of $3000$~K. The entire configuration is centred within a simulation box of $200\times200\times4000~{\rm kpc}^3$. We impose a spatially adaptive mass resolution in the CGM, ranging from $20~\msun$ near the galactic disc to $10^5~\msun$ in the outer regions. In addition, a tracer-based refinement strategy is implemented such that regions where more than 10\% of the gas mass originates from the disc are refined to the target mass of $20~\msun$.

We introduce an NSC at the location of the minimum gravitational potential of the galaxy, hosting a $10^4~\msun$ BH. Based on observational studies \citep{Georgiev+2016}, we adopt the NSC mass and effective radius reported for dwarf galaxies with stellar masses comparable to our model (see also Fig.~1 of \citetalias{Shin+2025}). In this study, the NSC has a total mass of $3.16\times10^5~\msun$ and an effective radius of $5$~pc. The NSC is modelled with a \citet{Hernquist1990} profile and is represented by $5\times10^5$ stellar particles of equal mass, representing old stars. These particles interact only gravitationally and do not undergo stellar evolution or feedback. The gravitational softening length for both the BH and NSC particles is set to $0.175$~pc.

\subsection{Gravity and hydrodynamics}
\label{sec2:code}

We perform our simulations using the gravito-hydrodynamics solver {\sc Arepo} \citep{Springel+2010, Pakmor+2016}, which computes gravitational forces via a hierarchical tree algorithm and solves hydrodynamics using a second-order finite-volume Godunov scheme. The fluid is evolved on an unstructured moving mesh generated through Voronoi tessellation, enabling fully adaptive spatial resolution. Barring a central region around the BH, mesh refinement follows a quasi-Lagrangian scheme that keeps gas cell masses close to a target value of $20~\msun$.

Radiative heating and cooling processes are modelled using the {\sc Grackle} chemistry and cooling library\footnote{\url{https://grackle.readthedocs.io/}} \citep{Smith+2017}, which tracks the non-equilibrium abundances of six primordial species: $\ion{H}{I}$, $\ion{H}{II}$, $\ion{He}{I}$, $\ion{He}{II}$, $\ion{He}{III}$, and free electrons. Metal-line cooling is included through tabulated rates computed with the photoionisation code {\sc Cloudy} \citep{Ferland+2013}. We also include a metagalactic UV background based on the \citet{Haardt+Madau2012} model, with self-shielding applied according to the prescription of \citet{Rahmati+2013}, as implemented in {\sc Grackle}.

\begin{figure*}
\centering
\includegraphics[width=\textwidth]{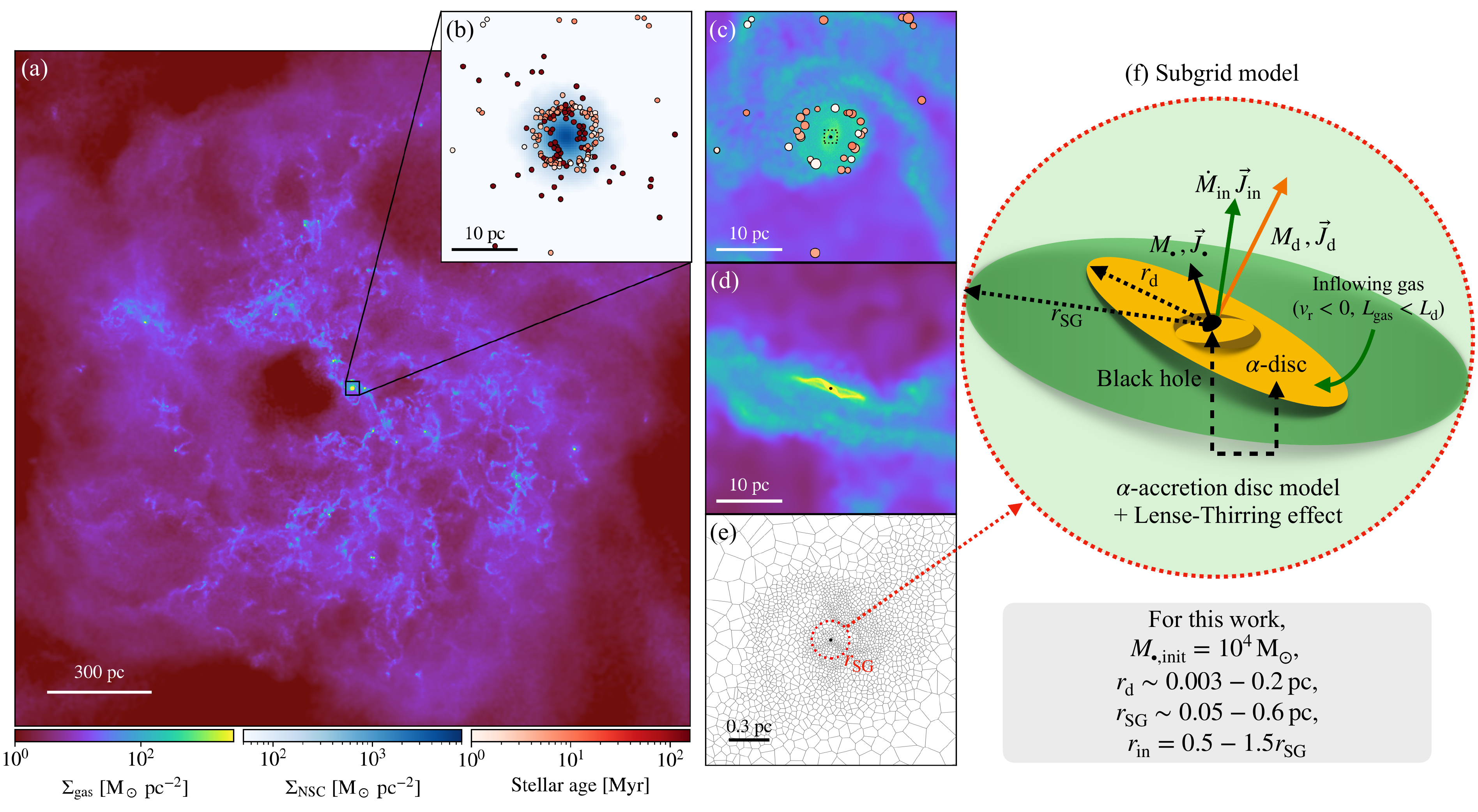}
\caption{Visualisation of our simulation setup. (a) Face-on gas column density projection of the {\tt full} run in a box of $2$~kpc on-a-side at $t=136$~Myr. (b) NSC surface density projection in a $40$~pc box. {\it Coloured dots} mark the positions of stars formed during the simulation; the dot colours correspond to stellar ages. (c) Zoom-in, face-on view of the surface gas density projection in a $40$~pc box. Due to the NSC potential, gas inflows effectively form a CND with a radius of approximately $6$~pc. The reddish dots indicate massive young stars ($M_\star > 5~\msun$, age $< 35~\mathrm{Myr}$) which are responsible for stellar feedback. The dot size scales with stellar mass, ranging from $5.12~\msun$ to $18.2~\msun$. The {\it black dot} marks the BH position. (d) Zoomed-in, edge-on view of the gas surface density projection in a $40$~pc box. The CND is warped as the angular momentum of the inflowing gas changes over time. (e) Voronoi slice plot of a $2$~pc box region illustrating the super-Lagrangian refinement, where gas cell sizes decrease toward the BH and achieve a spatial resolution of $<0.01$~pc. At the self-gravity radius, $r_{\rm SG}\sim0.15$~pc, the gas is resolved with $\sim2000$ cells. (f) Sketch of our subgrid BH accretion model. At $r_{\rm SG}$, we measure the inflow rates of mass and angular momentum, which are then used to update the $\alpha$-disc mass and angular momentum. Based on the $\alpha$-disc model \citep{Shakura+Sunyaev1973} and the Lense–Thirring effect \citep{Bardeen+Petterson1975}, the subgrid model computes the evolution of the BH mass, BH spin, accretion disc mass and angular momentum. See Section~\ref{sec:bh-model-method} for further details.}

\label{fig:main}
\end{figure*} 
\begin{table*}
\centering
\caption{List of simulations performed in this study. The parameters are: the stellar feedback model, star formation threshold mass, star formation efficiency, gas minimum softening length, metallicity of galactic disc, newly-created stellar mass in the galactic region, $M_{\rm new\,\star, total}$ ($r<1$~kpc), newly-created stellar mass in the nuclear region ($r< 10$~pc), $M_{\rm new\,\star, NSC}$, and the BH mass, $M_{\bullet}$, at $t=450$~Myr.}
\begin{tabular}{ccccccccc} %
\hline
Setup name & Stellar FB model & SF condition &$\epsilon_{\rm SF}$ &$\epsilon_{\rm soft,gas}$ &$Z$&$M_{\rm new\,\star, total}$ & $M_{\rm new\,\star, NSC}$& $M_{\bullet}$\\
  &&&&[pc] &[$Z_\odot$]&[10$^5\msun$]  &[$10^3\msun$]  &[$10^4\msun$]\\
\hline\hline
{\tt noFB}&- &$M_{\rm J}<N_{\rm J}m_{\rm cell}$&0.02&1.75&0.1&115&{303}&1.36\\
{\tt PIPE}& PI+PE&$M_{\rm J}<N_{\rm J}m_{\rm cell}$&0.02&1.75&0.1&16.3&{152}&1.46\\
{\tt SN}& SN&$M_{\rm J}<N_{\rm J}m_{\rm cell}$&0.02&1.75&0.1&4.01&4.33&1.15\\
{\tt full}& SN+PI+PE&$M_{\rm J}<N_{\rm J}m_{\rm cell}$&0.02&1.75&0.1&5.62&8.06&1.22\\
\hline
{\tt fixed-SFthr}& SN+PI+PE&$M_{\rm J}<N_{\rm J}\times20\msun$&0.02&1.75&0.1&5.43&7.36&1.19\\
&&+Convergent flow&&&&&&\\
{\tt SFeff100}& SN+PI+PE&$M_{\rm J}<N_{\rm J}m_{\rm cell}$&1.0&1.75&0.1&6.51&15.5&1.21\\
{\tt 1/2soft}& SN+PI+PE&$M_{\rm J}<N_{\rm J}m_{\rm cell}$&0.02&0.875&0.1&5.83&16.9&1.39\\
{\tt lowZ}& SN+PI+PE&$M_{\rm J}<N_{\rm J}m_{\rm cell}$&0.02&1.75&0.01&1.50&7.05&1.30\\
\hline
\end{tabular}
\label{tab:runs}
\end{table*}
\subsection{Stellar physics}
\label{sec2:SFFB}
\subsubsection{Star formation: explicit IMF sampling}

Our model includes star formation and the associated feedback from individually tracked massive stars, following the methodology of \citet{Smith+2021}. For a detailed description and validation of the implementation, we refer the reader to that work, and we briefly summarise the key elements relevant to this study below.

To model star formation, we first evaluate the Jeans stability of each gas cell by comparing its mass to the local Jeans mass, defined as
\begin{equation}
M_{\rm J} = \frac{\pi^{5/2} c_{\rm s}^3}{6 G^{3/2} \rho^{1/2}},
\end{equation}
where $c_{\rm s}$ is the sound speed, $G$ is the gravitational constant, and $\rho$ is the gas density. A gas cell is considered star-forming if $M_{\rm J} < N_{\rm J} m_{\rm cell}$, 
where the Jeans number $N_{\rm J} = 8$ controls the resolution of gravitational collapse.
Once a cell is identified as star-forming, its star formation rate (SFR) is computed as 
\begin{equation}
\dot{m}_{\star} = \epsilon_{\rm SF}\,{m_{\rm cell}}/{t_{\rm ff}},
\label{eq:sfeff}
\end{equation}
where $t_{\rm ff} = \sqrt{3\pi/(32 G \rho)}$ is the free-fall time, and $\epsilon_{\rm SF} = 0.02$ is the star formation efficiency, consistent with observed values in dense molecular regions \citep[e.g.,][]{Krumholz+Tan2007}. To ensure sufficient temporal resolution of the gas depletion process, we restrict the timestep of star-forming cells to be no larger than $0.1 m_{\rm cell} / \dot{m}_{\star}$, in addition to other integration constraints. Cells with non-zero SFRs are then stochastically converted into collisionless star particles in proportion to their computed SFRs. 
Then we assign individual stellar masses to each newly formed star particle by explicitly sampling from the \cite{Kroupa2001} initial mass function (IMF) over the range of $0.08$--$100~\msun$. These stars collectively represent the internal stellar content of the particle, and all feedback processes are directly coupled to the evolution of these individual stellar components\footnote{This approach differs from the conventional SSP approximation, in which each star particle is assumed to fully sample the IMF. Such simplified treatments fail to capture the small-scale spatio-temporal clustering of ionizing sources from OB stars, which significantly affects the ability of photoionisation feedback to regulate star formation in the ISM \citep[see][]{Smith2021}.}. 
For details on how we preserve the IMF shape while addressing discrepancies with the dynamical masses of individual star particles, we refer the reader to \citet{Smith2021}.

\subsubsection{Early stellar feedback: photoionisation (PI) \& photoelectric (PE) heating}
\label{sec:pre-SN}
As part of the early stellar (or pre-SN) feedback, we implement both photoionisation and photoelectric heating, sourced by the interstellar radiation field (ISRF) from massive stars. Feedback quantities are derived by interpolating lookup tables as a function of the zero-age main sequence (ZAMS) mass, based on the \citet{Kroupa2001} IMF. To reduce complexity, the stellar metallicity is fixed at $0.1~Z_\odot$ rather than interpolating across metallicity. Stellar lifetimes are taken from the PARSEC evolutionary tracks \citep{Bressan+2012}, while binary evolution is neglected. Emission rates of far-UV (FUV; $6$--$13.6$~eV) and extreme-UV (EUV or ionizing; $>13.6$~eV) photons are drawn from the OSTAR2002 stellar atmosphere models \citep{Lanz+2003, Emerick+2019}. For simplicity, these photon outputs are assumed to remain fixed at their ZAMS values throughout each star’s lifetime.

To track the spatial and temporal evolution of $\ion{H}{II}$ regions, we employ a directionally resolved Str\"omgren-type approximation. For each ionising star, the neighbouring volume is divided into 12 angular pixels, within which the local balance between ionising photon production and recombination rate is computed, accounting for overlapping contributions from other sources \citep[see][]{Smith+2021}. This approach avoids the density-weighted bias inherent to the spherical symmetry and provides a more accurate representation of the growth and structure of ionised regions over time. The maximum extent of each $\ion{H}{II}$ region is limited to 50~pc. Gas cells identified as lying within an $\ion{H}{II}$ region are heated to $10^4$~K and are prevented from cooling below this temperature for the duration of their ionised state. 

The ISRF in the $6$--$13.6$~eV range heats dust grains via the photoelectric effect. The dust-to-gas mass ratio is determined from the metallicity using the scaling relation of \citet{Remy-Ruyer+2014}. To estimate the FUV energy density at the location of each gas cell, we employ a simplified radiative transfer scheme that assumes local attenuation at both the emitting and receiving ends, while treating the intervening medium as optically thin. This approach allows flux contributions to be accumulated during the gravity tree walk, after which the resulting heating rate is passed to {\sc Grackle} as an additional source term.

\subsubsection{SN feedback}
Stars with ZAMS masses between $8$ and $35~\msun$ end their lives as core-collapse SNe, injecting mass, metals, energy, and momentum into the surrounding gas. Injection is applied to the host gas cell and its immediate neighbours, defined as cells sharing a face with the host. Feedback quantities are distributed to preserve isotropy. To account for unresolved momentum in the Sedov–Taylor phase of the SN remnant, we implement a mechanical feedback scheme; however, at the resolution used in this study, most SNe are sufficiently well resolved \citep{Smith+2021}. Type Ia SNe, stellar winds, and runaway OB stars are not included in our model. Each SN is assigned a fixed explosion energy of $10^{51}$~erg, with progenitor-mass-dependent ejecta masses and metallicities taken from \citet{Chieffi+2004}.

\subsection{Super-Lagrangian refinement around BHs}
\label{sec2:SL}
As in \citetalias{Shin+2025}, we adopt the super-Lagrangian refinement strategy introduced by \citet{Curtis+Sijacki2015} to accurately resolve the transport of mass and angular momentum from the ISM down to the self-gravity radius of the accretion disc and the BH system. In this scheme, a refinement region of radius $R_{\rm ref}$ is defined around the BH, within which the maximum allowed cell radius increases linearly with distance from the BH. Specifically, $R^{\rm (cell)}_{\rm max}(r)$ grows linearly from $0.01$~pc at the BH location to $0.8$~pc at the outer boundary of the refinement region, $R_{\rm ref} = 6$~pc.
Within this refinement volume, cells are split or merged such that their radii remain within a controlled range: ${R^{\rm (cell)}_{\rm max}(r)}/{C} < r_{\rm cell}(r) < R^{\rm (cell)}_{\rm max}(r)$,
where the refinement factor $C=4$ regulates the permitted span in cell sizes.
The innermost resolution limit, given by $R^{\rm (cell)}_{\rm max}(0)/C = 2.5 \times 10^{-3}$~pc, is sufficient to resolve the self-gravitating radius of the accretion disc, whose radius always exceeds $0.05$~pc in our BH system. \ej{Details of the spatial and mass resolution distributions can be found in Fig.~2 of \citetalias{Shin+2025}.}
To prevent the artificial formation of star particles with extremely low masses in this high-resolution region, star formation is explicitly suppressed in gas cells with masses below $0.08~\msun$.

Fig.~\ref{fig:main} highlights the multiscale nature of our simulations. On galactic scales (panel {\it (a)}), the gas distribution is highly clumpy and exhibits a multiphase, turbulent ISM structure, shaped by non-equilibrium cooling and stellar feedback. In the circumnuclear region (panels {\it(b)–(d)}), the NSC potential channels inflowing gas onto a CND of radius $\lesssim$7 pc around the central BH ({\it black dot}). In the innermost $2$~pc (panel {\it (e)}), the Voronoi mesh illustrates the super-Lagrangian refinement scheme, where gas cell sizes decrease toward the BH, reaching a spatial resolution better than $0.01$~pc and resolving the self-gravity radius of the accretion disc ($r_{\rm SG}$; {\it red circle}) with thousands of cells.

\subsection{Black hole accretion model}
\label{sec:bh-model-method}
\subsubsection{$\alpha$-accretion disc and black hole spin model}
We follow the method introduced in \cite{Fiacconi+2018} and \citetalias{Shin+2025}, adopting a steady-state $\alpha$-accretion disc model to describe BH accretion. Our model measures the mass and angular momentum flux from the surrounding ISM at the self-gravity radius of the accretion disc. Following \cite{Fiacconi+2018}, we analytically compute the evolution of the accretion disc and the BH according to the \citet{Shakura+Sunyaev1973} solution, incorporating the effect of Lense-Thirring precession on a viscous disc via the \citet{Bardeen+Petterson1975} effect. For further details of the implementation, we refer the reader to \citet{Fiacconi+2018} and \citetalias{Shin+2025}, and provide a brief summary here.

To estimate the mass flux toward the accretion disc, we evaluate gas cells within a spherical shell spanning $0.5~r_{\rm SG} < r < 1.5~r_{\rm SG}$, selecting only those cells with inward radial velocity ($v_{\rm r} < 0$) and that would circularize onto the accretion disc ($L<L_{\rm d}$). Here, $r$ is the radial distance from the BH, $v_{\rm r}$ the radial component, and $L$ and $L_{\rm d}$ are the specific angular momenta of the gas cell and the $\alpha$-accretion disc, respectively. The inflow rate, $\dot{M}_{\rm in}$, and associated specific angular momentum, $\vec{L}_{\rm in}$, are then computed as:
\begin{equation}
\dot{M}_{\rm in}=
\frac{\sum m_i v_{{\rm r}, i}}{\Delta x_{\rm shell}}\,,
\end{equation}
\begin{equation}
\vec{L}_{\rm in}=
\frac{\sum m_i v_{{\rm r}, i}\vec{L}_{i}}{\sum m_i v_{{\rm r}, i}}\,,
\end{equation}
where $m_i$, $v_{{\rm r}, i}$, and $L_i$ are the mass, radial velocity (relative to the BH), and specific angular momentum of the $i$-th gas cell. The summation is performed over all gas cells within a shell of width $\Delta x_{\rm shell} = r_{\rm SG}$ that satisfy the inflow and circularisation criteria. The inflowing specific angular momentum, $\vec{L}_{\rm in}$, is a mass-flux-weighted average of the angular momenta of the inflowing gas. The quantities $\dot{M}_{\rm in}$ and $\vec{L}_{\rm in}$ are then used to update the $\alpha$-disc according to
$\dot{M}_{\rm d}=\dot{M}_{\rm in}-\dot{M}_{\bullet,0}$ and 
$\dot{\vec{J}}_{\rm d}=\dot{M}_{\rm in}{\vec{L}}_{\rm in}-\dot{\vec{J}}_{\bullet,0}$, where $\dot{M}_{\rm d}$, $\dot{\vec{J}}_{\rm d}$ are the mass and angular momentum inflow rate to the accretion disc, and $\dot{M}_{\bullet,0}$, $\dot{\vec{J}}_{\bullet,0}$ denote the respective inflows from the accretion disc to the BH.

The BH and accretion disc are initialised with masses of $10^4\msun$ and $10^2\msun$, respectively, and with the Eddington fraction of $f^{(\rm init)}_{\rm Edd} = 0.003$. The initial spin parameter of the BH is set to $a_{\bullet}=0.7$. These parameters uniquely determine the initial angular momenta of both the BH and the accretion disc. We initialise the angular momentum vectors of the BH and the disc to be aligned with that of the galactic disc. \ej{Note that in the present study we do not include BH feedback physics, which will be studied in detail in our forthcoming work.}

\begin{figure*}
\includegraphics[width=\textwidth]{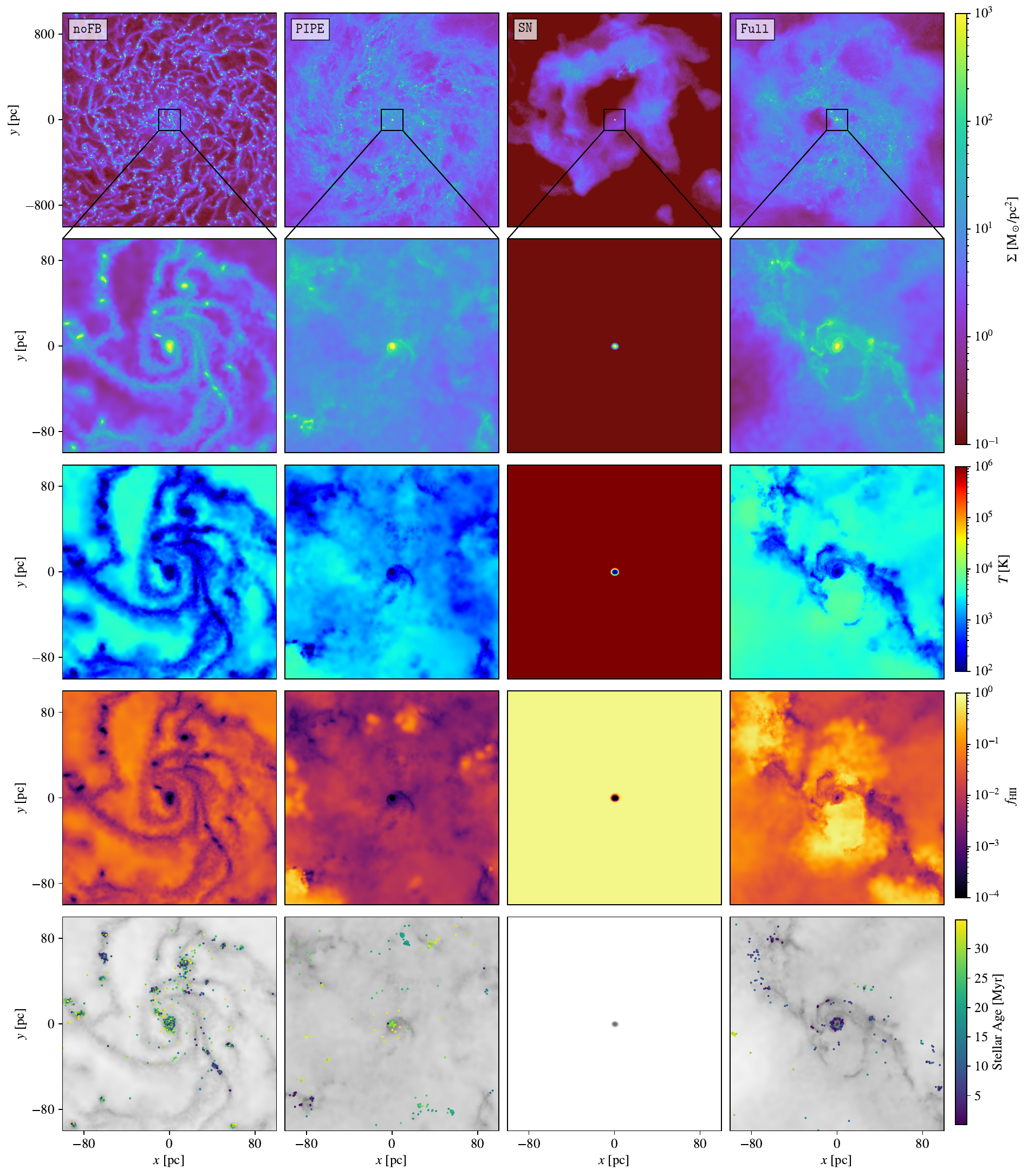}
    \caption{Face-on projections of the galactic disc. Simulation results at $t=140$~Myr for four different stellar feedback schemes. {\it Top row}: Gas column density in a $2$~kpc box. {\it Second row}: Zoomed-in view of the gas column density in a $200$~pc box. {\it Third row}: Density-weighted temperature. {\it Fourth-rows}: $\ion{H}{II}$ fraction. {\it Bottom row}: Locations of young stars overlaid on the gas column density background. The {\it dot colours} indicate stellar ages. The thermodynamic state and morphology of the ISM are strongly affected by the stellar feedback combination, but all simulations consistently form a cold dense CND within $\lesssim 10$~pc at the centre. See Section~\ref{sec3:gas-proj} for further details.}
\label{fig:proj}
\end{figure*}
\subsection{Suite of simulations}
\label{sec:overall-setups}
Building on the setup introduced in the previous section, we examine how different numerical and physical configurations affect the evolution of the nuclear region and the central BH over a simulation time of 450~Myr. The simulations are divided into two main groups, summarised in Table~\ref{tab:runs} along with their set-ups and results. The first group explores different stellar feedback channels, including different combinations of early stellar feedback (PI, PE), and SN processes. This design enables us to isolate the individual contributions of radiation and SN feedback on the properties of the CND and BH growth.

The second group builds on our fiducial configuration ({\tt full}) but systematically varies the star formation criteria, the minimum gas softening length, and the initial gas metallicity. In the {\tt fixed-SFthr} run, the star formation criterion is set relative to the target gas mass rather than the local cell mass, i.e. $M_{\rm J}<N_{\rm J}\,m_{\rm gas, target}$\footnote{In the super-Lagrangian refinement region, the gas cell mass decreases with decreasing distance from the BH. In this test, we fix the star formation threshold so that stars form under the same threshold conditions as outside of the super-Lagrangian region.}. We also imposed a strict converging-velocity criterion, requiring each velocity component to satisfy $\partial v_i / \partial x_i < 0$, for $i = x, y, z$. We adopted this criterion due to the presence of strong shear flows in the CND and accreting gas, where the Jeans mass alone is not always a reliable predictor of fragmentation, to gauge the robustness of our results. In the {\tt SFeff100}, we set $\epsilon_{\rm SF}=1$ in Equation~\ref{eq:sfeff}, instead of the fiducial value of $0.02$. In the {\tt 1/2soft} run, the gas softening length is reduced by a factor of two. Finally, in the {\tt lowZ} run, the initial metallicity of the gaseous galactic disc is a tenth of the fiducial value (i.e. $Z=0.01\,Z_\odot$), enabling us to explore how a more pristine galactic environment influences the CND and MBH accretion\footnote{In the {\tt lowZ} run, we kept the stellar metallicity at $0.1\,Z_\odot$, the same as in the other runs, to enable a fair comparison. Note that in our models the ionising photon production rate is not very sensitive on the stellar metallicity.}.

 \begin{figure*}
 \centering
\includegraphics[width=\textwidth]{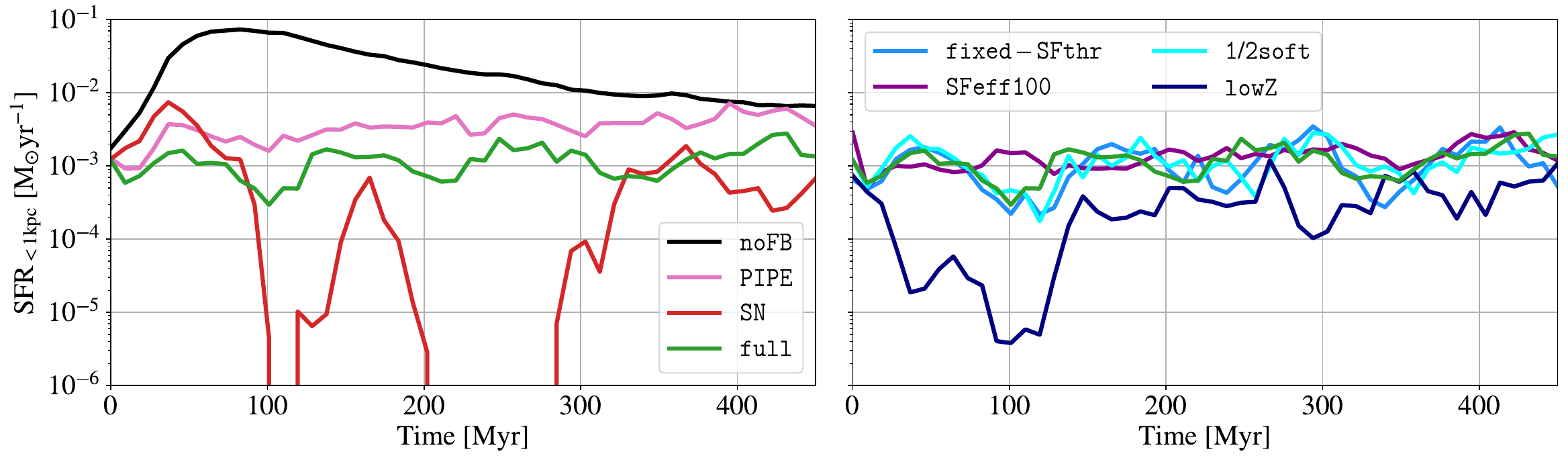}
    \caption{The time evolution of SFR of simulations within the $r=1$ kpc region for our simulations. The {\it left-hand} panel shows that the early stellar feedback leads to much smoother SFRs due to the suppression of excessive disc fragmentation, while the additional inclusion of SN feedback  (the {\tt full} model) introduces slightly burstier SFR evolution. The {\it right-hand} panel shows that our SFR evolution is robust against reasonable changes in our star formation prescription, while, as expected, lower initial gas metallicities affect gas cooling efficiency and hence the SFR. See Section~\ref{sec:sfr} for further details.}
    \label{fig:totSFR}
\end{figure*}

\section{Results}
\label{sec3:results}

\subsection{Simulation overview}
\label{sec3:gas-proj}

Fig.~\ref{fig:proj} presents face-on maps at $t=140$~Myr for simulations with different stellar feedback implementations. From top to bottom, the panels show the gas surface density, density-weighted temperature, $\ion{H}{II}$ fraction, and the spatial distribution of young stars (age $<35$~Myr), overlaid on a projected gas density background. The {\it top row} covers a $2$~kpc-a-side region, while the {\it bottom four rows} zoom in on the central $200$~pc of the galaxy. Different stellar feedback models result in markedly different morphologies in the galactic disc. However, importantly, all simulations consistently produce a compact, cold CND of size $\lesssim 10$~pc at the centre. As detailed in \citetalias{Shin+2025}, the CNDs form as ISM gas loses angular momentum through orbit crossing under the gravitational potential of the NSC. The CND gas subsequently circularises into an accretion disc with a characteristic scale of 0.1~pc as it orbits within the (largely) isotropic NSC, transferring angular momentum to the cluster (see \citetalias{Shin+2025} for a detailed analysis).  

Unsurprisingly, in the {\tt noFB} run, the gas disc appears highly fragmented, with dense clumps hosting active star formation. A notable fraction of stars also form near the CND. However, in the absence of stellar feedback, the $\ion{H}{II}$ fraction remains low throughout the ISM. In contrast, the {\tt PIPE} run, which includes only early stellar feedback, shows significantly less fragmentation. Comparing the $\ion{H}{II}$ fraction and young stellar distribution panels confirms that ionised regions are closely aligned with the locations of recently formed stars. In the zoom-in panels, we find that the younger stars tend to be associated with relatively diffuse gas, whereas stars older than $\sim 20$~Myr are spatially decoupled from dense gas clumps. This demonstrates that, while stars initially form in high-density environments, massive stars ($>5~\msun$) rapidly ionise and heat the surrounding gas through radiation feedback, hence efficiently disrupting their natal clumps.

In the simulation with only SN feedback ({\tt SN}), the ISM exhibits the lowest densities and highest temperatures among all models. This is consistent with the findings of \citet{Smith+2021}, which reported that in the absence of early stellar feedback, the disc becomes highly fragmented. As a result, SNe tend to be clustered and inject energy in a strongly bursty manner. Following such bursty events, once the feedback cycle ends, the ISM becomes diffuse and hot, remaining in this state until it cools again via radiative processes on a $\sim300$~Myr timescale. As we will further discuss in a later section, the CND survives this in-situ star formation-driven SN activity. However, after the SN episodes subside, the CND becomes Toomre-stable, ceases in-situ star formation, and begins to feed the central BH instead.

Lastly, in the simulation that includes full stellar feedback physics, the ISM structure is qualitatively similar to that of the {\tt PIPE} run in terms of dense gas distribution. However, it also exhibits large voids of $300$--$500$~pc in size, similar to those seen in the {\tt SN} run. As in the {\tt PIPE} run, the $\ion{H}{II}$ fraction is high in regions that spatially trace young stars. These stars are not strongly clustered, with only a few forming recognisable clusters, owing to the `dispersive' effect of early stellar feedback. Importantly, however, there is a central concentration of (young) stars which are born in the CND.

\subsection{Global SFRs}
\label{sec:sfr}

Fig.~\ref{fig:totSFR} shows the star formation histories within a $1$~kpc radius from the galactic centre for all simulated galaxies examined in this study. The simulation setups and the total stellar mass formed in each run are summarised in Table~\ref{tab:runs}. Focusing first on the {\it left} panel, which compares simulations with different stellar feedback models, it is evident that the star formation history is highly sensitive to the adopted feedback prescription, consistent with the findings of \cite{Smith+2021}. In the absence of any stellar feedback ({\tt noFB}), the disc begins to fragment significantly, resulting in a rapid rise in the SFR that peaks at approximately $0.1~\msun\,{\rm yr}^{-1}$. As the gas reservoir is gradually depleted, the SFR steadily declines, reaching $\sim 0.01~\msun\,{\rm yr}^{-1}$ by $350$~Myr. A similarly sharp increase in the SFR is observed in the simulation including only SN feedback ({\tt SN}) during the first $30$~Myr. This occurs because, without early stellar feedback and with the SN explosion delay time of 3--100~Myr, the disc undergoes excessive fragmentation, similarly to the {\tt noFB} run. Consequently, in the {\tt SN} run, gas clumps form efficiently, leading to strongly clustered SN explosions and a markedly bursty star formation history. During $t=200$--$300$~Myr, this clustering of SN events effectively quenches star formation across the disc. 

By contrast, in the simulations incorporating early stellar feedback ({\tt PIPE} and {\tt full}), the SFR is immediately suppressed to $10^{-3}~\msun\,{\rm yr}^{-1}$ as a result of photoionisation effects once stars begin to form. In the {\tt PIPE} run, the average SFR remains at $\sim3\times10^{-3}~\msun\,{\rm yr}^{-1}$, with the lowest level of burstiness among all models. Finally, in the simulation including all feedback processes, early stellar feedback suppresses disc fragmentation more effectively than in the SN-only case, resulting in an SFR lower by a factor of 2--3 while showing slightly greater burstiness.

Next, the {\it right} panel of Fig.~\ref{fig:totSFR} compares how the SFR varies as a function of time when adopting different choices of star formation prescriptions, gas softenings, and initial gas metallicity compared to the {\tt full} run. With the exception of the {\tt lowZ} simulation, reassuringly, all runs exhibit broadly similar SFRs to the {\tt full} run. In the {\tt lowZ} case, the lower metallicity reduces cooling efficiency, resulting in a newly-formed stellar mass that is about $30$~\% of that in the {\tt full} run (see Table~\ref{tab:runs}). All simulations also show a significant decline in SFR during the first $20$~Myr, which can be attributed to the photoionisation feedback, as discussed earlier. Among them, the {\tt eff100} run exhibits the most intense star formation activity in the early phase, as expected, but the SFR subsequently drops to a similar level as in the other simulations, reaching $\sim10^{-3}\,\msun\,{\rm yr}^{-1}$. These results indicate that, at this resolution ($m_{\rm gas, target}=20\msun$), star formation becomes self-regulated by feedback, largely independent of the adopted star formation efficiency. \ej{In future work it will be important to quantify the impact of BH feedback on star formation as well.}

\begin{figure*}
\includegraphics[width=\textwidth]{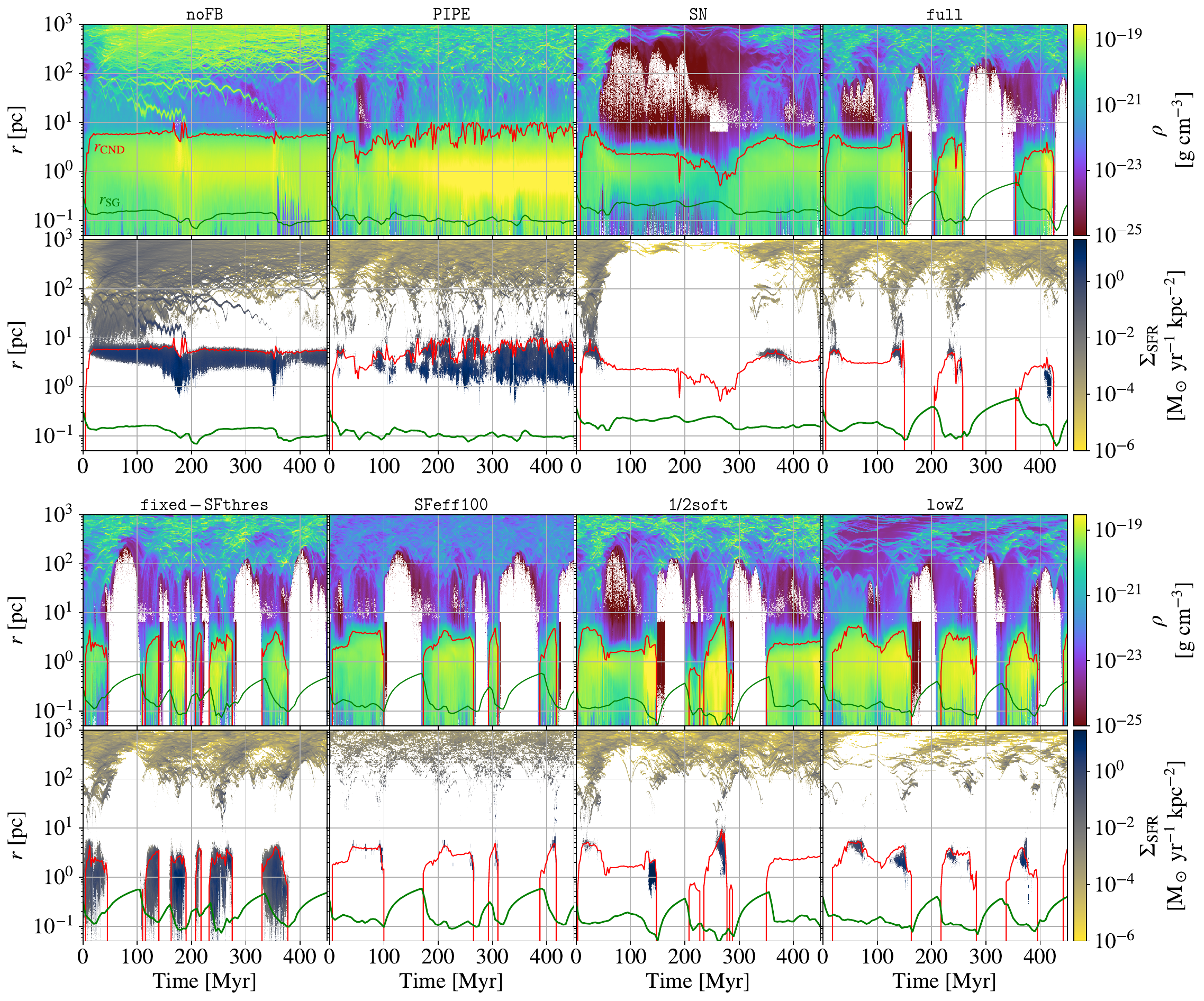}
    \caption{Time evolution of the radial profiles of gas density (density-weighted) and SFR surface density. The {\it red} lines show the evolution of the CND radius, while the {\it green} lines mark the evolution of the self-gravity radius of the accretion disc. The evolution of the ISM, the CND properties and the spatial distribution of SFRs vary with different stellar feedback channels and numerical setups explored. In all simulations that incorporate the combined effect of stellar radiation and SN feedback the CND is periodically disrupted and exhibits intermittent evolution. See Section~\ref{sec:3d} for further details.}
    \label{fig:3d}
\end{figure*}

\subsection{Time evolution of the multi-phase ISM}
\label{sec:3d}
Fig.~\ref{fig:3d} shows the time evolution of the azimuthally averaged radial profiles of density-weighted gas density and SFR surface density. The $y$-axis denotes the distance from the BH. The {\it red} lines trace the time evolution of the CND radius, which we define as the radial location where both the density exceeds $10^{-20}{\rm g\,cm}^{-3}$ and the temperature falls below 30~K. If no radius satisfies the density criterion, we instead adopt the radius that satisfies only the temperature condition. The {\it green} lines indicate the self-gravity radius of the accretion disc. The time-averaged profiles of various key properties are provided in Appendix~\ref{sec:appendix-prof}. 

First, we consider the upper panels, which compare the results of different stellar feedback models. As already shown in Fig.~\ref{fig:proj}, the structure of the galactic ISM varies significantly depending on the adopted stellar feedback schemes. In general, the gas density is relatively high in the $100~{\rm pc}$--$1$~kpc region, but it drops sharply at $r \sim 10$--$20$~pc. This decrease arises because the strong tidal field of the NSC stretches dense gas clumps and inflowing streams, reducing the local density relative to the surrounding medium. At radii corresponding to the CND ($r\sim1$--$10$~pc), the density significantly rises again, but within the self-gravity radius of the accretion disc, it decreases to $\sim 10^{-21}{\rm g\,cm^{-3}}$ or lower due to BH accretion.

As expected, the {\tt noFB} run exhibits high gas densities in the $100$~pc–$1$~kpc region. In the absence of feedback, once a gas clump forms, it either converts all of its gas into stars or continues to accrete material, thereby surviving for a very long time. Some of these gas+star clumps are drawn into the NSC potential and eventually merge with the CND+NSC system (see instances at $t=180$ and $350$~Myr). When such clumps collide with the CND+NSC, the CND experiences strong disturbances lasting for $\sim 30$~Myr, temporarily increasing its density. At other times, however, the CND radius and density remain nearly constant. Regarding the global SFR evolution, as shown earlier in Fig.~\ref{fig:totSFR}, the initial SFR is very high but gradually decreases as the gas mass is consumed. Central star formation is generally concentrated in the $r=3$--$10$~pc region, but during the merger events with external clumps at $t=180$ and $350$~Myr, star formation also occurs within the inner $1$~pc.

In the {\tt PIPE} run, the gas density across the $r=100~{\rm pc}$--$1$~kpc range is significantly lower than in the {\tt noFB} case due to the early stellar feedback. Interestingly, however, the peak density of the CND is the highest among all feedback models, and its radius remains larger than in the other simulations. This occurs because the radiative feedback suppresses star formation in the ISM, leaving more gas unconsumed by stars and thus available to accrete onto the CND. Within the CND, the SFR surface density is initially modest but gradually increases as the CND density builds up. The SFR surface density is highest at $0.4~{\rm pc}<r<3$~pc, whereas the gas density peaks at smaller radii, indicating that feedback compresses the gas inward. Stellar feedback also triggers short, intermittent episodes of star formation within the inner $\sim1$~pc. 

In the {\tt SN} run, the overall gas density is the lowest among all simulations. In particular, after the initial starburst, the gas density outside the CND radius drops below $10^{-24}~{\rm g\,cm^{-3}}$, and the galactic disc becomes nearly quenched. As shown earlier in Fig.~\ref{fig:proj}, when only SN feedback is included, clustered SNe heat the ISM very efficiently, suppressing both gas cooling and inflow into the CND, while gas draining onto the $\alpha$-disc continues. As a result, the CND mass gradually decreases, and its density remains the lowest among all simulations. \ej{By $t \sim 300$~Myr, however, the heated gas, much of which remains gravitationally bound due to confinement by the CGM pressure (see also Appendix~A of \citetalias{Shin+2025}), eventually undergoes radiative cooling in the central region. At late times, the cooling time becomes shorter than the dynamical time, allowing the gas to lose energy and fall back toward the centre. This leads to the gradual replenishment of the inner ISM and enables renewed star formation as well as renewed gas inflow onto the CND.}

In the {\tt full} run, the gas density in the $100~{\rm pc}$--$1$~kpc region lies between that of the {\tt PIPE} and {\tt SN} runs. Relative to {\tt PIPE}, this highlights the efficiency of SN feedback in heating the ISM, while compared to {\tt SN}, the {\tt full} run shows more frequent inflow episodes ($\lesssim100$~Myr). A distinctive feature of the {\tt full} run is that the CND forms only episodically, a behaviour also seen in other setups that include all feedback processes. In each CND episode, the CND typically survives for as short as 10~Myr and as long as $\sim100$~Myr before being disrupted by in-situ stellar feedback. Star formation in the circumnuclear region first appears near the CND radius but progressively shifts inward, as also shown in \citetalias{Shin+2025}. This is caused by stellar feedback compressing the gas inside the CND and enhancing its density.

Next, we consider the lower panels, which show simulations that adopt the same stellar feedback combination as the {\tt full} run while varying star formation, gas softening and initial gas metallicity choices. As expected, they show qualitatively similar behaviour to the {\tt full} run: {\it (1)} episodic star formation in the central $0.1$--$100$~pc region, {\it (2)} $4$--$8$ inflow episodes over a time period of 450~Myr, and {\it (3)} the CND existing only episodically, with lifetimes ranging from as short as $\sim10$~Myr to more than $100$~Myr. In most runs, star formation in the inner pc region is suppressed because, due to the super-Lagrangian refinement, the cell mass decreases with radius and only a few cells satisfy the Jeans criterion ($M_{\rm J} < 8\,m_{\rm cell}$). In the {\tt fixed-SFthr} run, however, star formation is allowed under conditions similar to those outside the refinement region, enabling stars to form even within the inner 0.1~pc of the CND\footnote{\ej{Furthermore, note that the innermost region around the BH should be gravitationally stable due to the strong potential of the BH and NSC, which leads to a high epicyclic frequency and correspondingly large Toomre parameter ($Q_{\rm Toomre} > 10$; see Fig.~9 in \citetalias{Shin+2025}). Star formation in this region is therefore expected to be strongly suppressed even without the SF restriction.}}. Interestingly, despite this more generous star formation criterion, the cumulative stellar mass in the CND is not significantly larger, because the feedback from in-situ stars self-regulates star formation within the CND itself (see Table~\ref{tab:runs}).

In the {\tt SFeff100} run, star-forming cells convert gas into stars very efficiently, causing dense structures to be disrupted rapidly. As a result, the ISM density in the $r=10~{\rm pc}$--$1$~kpc region is comparatively low. Interestingly, although the total stellar mass formed is slightly higher, it remains comparable within a factor of two to that in the other runs \citep[see also][]{Smith+2021}. Both the {\tt 1/2soft} and {\tt lowZ} runs show relatively enhanced CND densities. In the {\tt 1/2soft} run, the smaller gas gravitational softening allows structures to gravitationally collapse to smaller scales, leading to higher densities. In the {\tt lowZ} run, the reduced metallicity lowers cooling efficiency, thereby reducing the SFR (see Fig.~\ref{fig:totSFR}) and allowing more gas to accrete onto the CND.

\begin{figure*}
\centering
\includegraphics[width=0.955\textwidth]{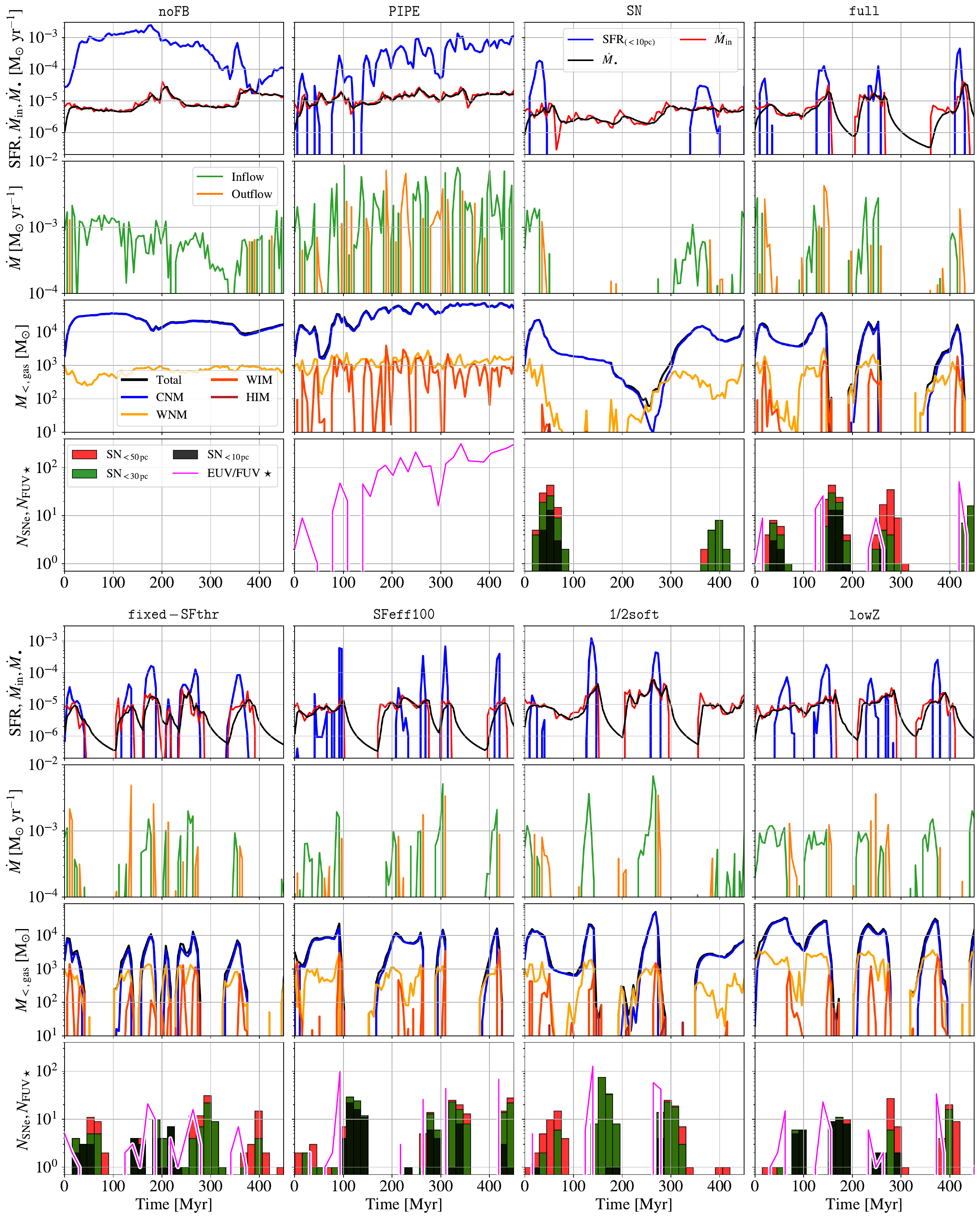}
    \caption{Evolution of CND gas and stars. {\it Top} and {\it fifth rows}: SFR within 10~pc ({\it blue lines}), mass inflow rate into the accretion disc ($\dot{M}_{\rm in}$; {\it red lines}), and BH accretion rate ($\dot{M}_{\bullet}$; {\it black lines}). {\it Second} and {\it sixth rows}: inflow ({\it green lines}) and outflow ({\it orange lines}) rate at $r=10$~pc. {\it Third} and {\it seventh rows}: enclosed gas mass within the central $10$~pc. The total gas mass is shown in black, and we also show gas mass split up in different phases, where the definitions of HIM, WIM, WNM, and CNM are provided in Section~\ref{sec:CNDcycle}. {\it Forth} and {\it bottom rows}: number of SNe events ($N_{\rm SNe}$; {\it red, green, black bars}) within three different radii and massive stars producing EUV/FUV photons within 10~pc ($N_{\rm FUV\star}$; {\it magenta lines}) formed within each time bin ($\Delta t=15$~Myr). EUV/FUV feedback prevents excessive fragmentation of the ISM, allowing more gas to flow into the circumnuclear region, while SN feedback suppresses fragmentation within the CND, making in-situ star formation episodic. Only when both feedback processes operate together does the CND undergo cyclic formation and destruction, leading to episodic accretion onto the BH accretion disc. See Section~\ref{sec:CNDsfr-BHAR} for details.}
    \label{fig:CNDcycle}
\end{figure*} 

\subsection{CND cycles: impact of the in-situ stellar feedback}
\label{sec:CNDcycle}

\subsubsection{In-situ star formation in the CND versus BH accretion rate}
\label{sec:CNDsfr-BHAR}

In this section, we quantitatively examine how the galactic nuclear region is affected by various feedback and modelling choices. The top panels of Fig.~\ref{fig:CNDcycle} show the time evolution of key physical quantities within a $10$~pc region around the BH for the four stellar feedback models: the SFR ({\it blue} lines), the mass inflow rate toward the accretion disc ($\dot{M}_{\rm in}$; {\it red} lines), and the BH accretion rate ($\dot{M}_{\bullet}$; {\it black lines}). The panel below shows the gas mass within the same $10$~pc region as a function of time. The {\it black} lines indicate the total gas mass, while the coloured lines represent the mass of gas in different thermal phases: (1) cold neutral medium (CNM): $T<500$~K (2) warm neutral medium (WNM): $500~{\mathrm K} <T< 5\times10^5$~K and $\chi_{\ion{H}{I}}>0.5$ (3) warm ionized medium (WIM): $500~{\mathrm K} <T<5\times10^5$~K and $\chi_{\ion{H}{I}}<0.5$ and (4) hot ionized medium (HIM): $T> 5\times10^5$~K. The bottom panel shows the number of massive stars ($M_\star > 5~\msun$; {\it magenta} lines) that produce EUV/FUV photons, along with the number of SN events ({\it bars}). The {\it bar} colours indicate the distance from the BH at which the explosions occur: {\it black}, {\it green}, and {\it red} correspond to events within 10~pc, 30~pc, and 50~pc, respectively.

The SFR within $r<10$~pc is highly sensitive to the adopted stellar feedback model. In the absence of SN feedback, namely, in the {\tt noFB} and {\tt PIPE} runs, star formation proceeds much more steadily than in other models. In {\tt PIPE}, however, the evolution is not fully continuous: during the first 150~Myr, EUV/FUV radiation feedback drives episodic bursts, and after sufficient gas has built up in the nuclear region, the SFR undergoes quasi-periodic oscillations between $10^{-4}$ and $10^{-3}\msun\,{\rm yr}^{-1}$ on timescales of several tens of Myr. These two models produce more than an order of magnitude higher stellar mass in the CND compared to the other runs (see Table~\ref{tab:runs}). Interestingly, after $\sim200$~Myr, the nuclear SFR in {\tt noFB} declines, reaching below $10^{-4}\msun\,{\rm yr}^{-1}$ except for a brief enhancement around $\sim350$~Myr caused by a clump merger. This decline arises because, without feedback, the galactic disc gas is excessively consumed by star formation (or locked into dense star-forming clumps), reducing the supply of gas available for accretion onto the nucleus.

In the remaining six simulations with SN feedback, the SFR within $r<10$~pc is episodic. As discussed earlier, SN feedback strongly heats the CND, rendering the gas Toomre-stable (see Fig.~9 in \citetalias{Shin+2025}, for the detailed discussion). As shown in the previous section, the main difference between the {SN-only} and {full-feedback} runs lies in the typical time interval between successive star formation episodes. Among the five simulations with full-feedback physics, all show $4-10$ star formation episodes over 450~Myr, each lasting only $10$--$50$~Myr. In particular, the {\tt SFeff100} and {\tt 1/2soft} runs exhibit SFR peaks approaching $10^{-3}~\msun~\mathrm{yr}^{-1}$ with relatively short durations, which are driven by the high star formation efficiency adopted in {\tt SFeff100} and by the enhanced gas density in {\tt 1/2soft} due to stronger gravitational interactions resulting from the smaller softening length.

Next, we examine the inflow and outflow rates measured within a 10~pc radius across different simulations. The time evolution of the inflow from the galactic disc into the circumnuclear region and the outflow driven by in-situ stellar feedback provides crucial insight into the lifetime and cyclic nature of the CND. In the absence of feedback, except for the intermittent interactions between external clumps and the NSC+CND system, inflows dominate throughout most of the simulated time, typically maintaining rates of $10^{-4}$--$10^{-3}~\msun~\mathrm{yr^{-1}}$. Most of the inflowing gas is consumed by star formation and BH accretion, keeping the total CND mass relatively steady at $(1-3)\times10^{4}~\msun$ without significant growth or depletion. 

When early stellar feedback is included, outflows driven by in-situ stars emerge, resulting in both inflow and outflow rates reaching the highest levels among all simulations, typically $10^{-3}$--$10^{-2}~\msun~\mathrm{yr^{-1}}$. 
This occurs because the in-situ SFR within the CND is also the highest in these runs. 
In the remaining six simulations that include SN feedback, inflow episodes become intermittent. In the {\tt SN} run, inflow occurs at rates of $\lesssim10^{-3}~\msun~\mathrm{yr^{-1}}$, while when early stellar feedback is included as well, heating by EUV/FUV radiation from in-situ stars leads to shorter yet more frequent, bursty inflows. An interesting finding is that in all six simulations with SN feedback, the timing of outflow peaks coincides with brief but strong boosts in the BH accretion rate (as discussed later). This indicates that SN feedback does not merely disrupt the CND and quench accretion, but can also momentarily enhance BH growth through rapid, short-lived inflow events.

Next, we examine the evolution of the gas mass within $r<10$~pc. 
Overall, the CND mass oscillates between $10^3\msun$ and $5\times10^4\msun$, with more than 90\% of the gas remaining in the CNM throughout most of the simulation time. 
In the {\tt noFB} run, the gas is dominated by CNM and WNM, and no ionised component is present. 
When early stellar feedback is included, a WIM phase appears, with the mass fractions of WIM and WNM becoming nearly comparable. 
In the runs with SN feedback, the HIM emerges intermittently around $50$--$70$~Myr, driven by in-situ SNe. 
Comparing the {\tt SN} and {\tt full} runs, the latter shows an enhanced fraction of warm gas when EUV/FUV-emitting stars are formed, particularly through the excitation of the WIM phase. 
Interestingly, the model with the largest CND mass is not {\tt noFB} but {\tt PIPE}. 
As shown in Fig.~\ref{fig:3d}, early stellar feedback regulates star formation, allowing more gas to accrete onto the CND. 
Similarly, in the {\tt lowZ} run, the average CND mass remains relatively high. 
The reduced metallicity leads to less efficient cooling, which suppresses star formation and consequently allows more gas to accumulate in the nuclear region. 
In general, star formation tends to begin once the gas mass reaches $\sim10^4\msun$. 
However, in the {\tt fixed-SFthr} run, star formation occurs even when the CND mass is lower. 
Nevertheless, stellar feedback still regulates nuclear star formation, keeping the total stellar mass formed comparable to that in the other simulations (see Table~\ref{tab:runs}).

\begin{figure*}
\includegraphics[width=\textwidth]{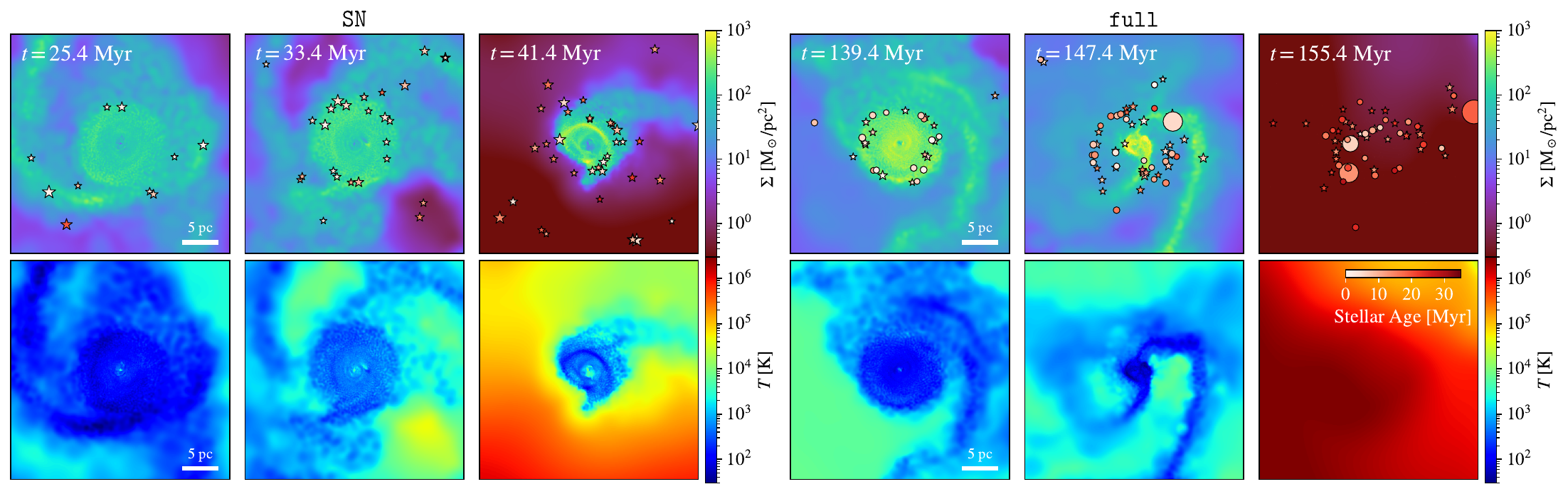}
    \caption{CND evolution in the {\tt SN} and the {\tt full} run. The panels show the gas surface density ({\it top} rows) and density-weighted temperature ({\it bottom} rows) projections at three stages: (1) 8~Myr before the onset of the first SNe in a given star formation generation, (2) immediately following the first SNe, and (3) 8~Myr afterwards. The projection box has a length of $30$~pc and is aligned with the angular momentum of the gaseous disc. {\it Circles} indicate the positions of massive stars that produce only radiative feedback, while {\it stars} denote those that produce SNe ($8\msun<M_{\rm \star, init}<35\msun$). The colours of the symbols represent stellar ages, and their sizes scale with stellar mass, ranging from $5.07\msun$ to $89.4\msun$. In the {\tt SN} run, the CND and its surrounding region remain very cold before the onset of SNe, whereas in the {\tt full} run, EUV/FUV radiation from massive stars heats and diffuses the CND, making it more vulnerable to subsequent SN feedback. See Section~\ref{sec:SN-full} for further details.}
    \label{fig:SN-full}
\end{figure*}

We now examine the inflow rate onto the accretion disc and the evolution of the BH accretion rate, which fall into two qualitative categories: runs with full-feedback physics and those without. As discussed earlier, in the full-feedback runs, the CND exists only episodically, so the inflow toward the accretion disc also appears intermittently. For example, the {\tt full} runs undergo two extended quiescent phases, at $t=150-200$~Myr and again at $t=250-410$~Myr. During these intervals, the CND is completely evaporated by SN feedback, and BH accretion continues solely from the accretion disc.

Examining the $\dot{M}_{\rm in}$ and BH accretion rate in more detail in the absence of feedback, the simulations show only small-amplitude, rapid oscillations in both $\dot{M}_{\rm in}$ and BH accretion rate\footnote{These rapid oscillations reflect the BH wobbling around the centre of the CND within a radius of $\sim0.1$~pc.}, except at $t\sim180$ and $t\sim350$~Myr when clump-CND mergers (see Fig.~\ref{fig:3d}) induce angular momentum transport and temporarily enhance gas inflow. 
Interestingly, the mass inflow rate onto the accretion disc in the {\tt PIPE} run is slightly higher than in the {\tt noFB} run. Following each SFR peak, the BH accretion rate shows a brief but distinct rise within $\sim10$~Myr, caused by feedback-driven compression of gas that enhances the density inside $r_{\rm SG}$\footnote{More precisely, feedback briefly expels a significant amount of gas from the central region near the BH. This heated gas then falls back under the influence of the NSC potential on a free-fall timescale of $<0.3$~Myr, leading to an enhanced density in the $r_{\rm SG}$ region.}. In addition, because early star formation is inefficient in the {\tt PIPE} run, the CND retains more mass than in {\tt noFB}. This larger reservoir, combined with the effects of stellar feedback, leads to an overall higher BH accretion rate.

It is noteworthy that in the {\tt SN} run, the number of SNe per episode is comparable to that in the {\tt full} run, yet the CND survives. This contrast suggests that early stellar feedback associated with in-situ star formation weakly heats and diffuses the CND, rendering it more susceptible to subsequent SNe (as discussed later). Thus, while the combination of early and SN feedback leads to the complete disruption of the CND, SN feedback alone is insufficient to produce the same effect. Nevertheless, despite the episodic nature of accretion in the {\tt full} run, the efficient inflow from the ISM enables more substantial BH growth than in the {\tt SN} run (as discussed later).

\begin{figure*}
\centering
\includegraphics[width=\textwidth]{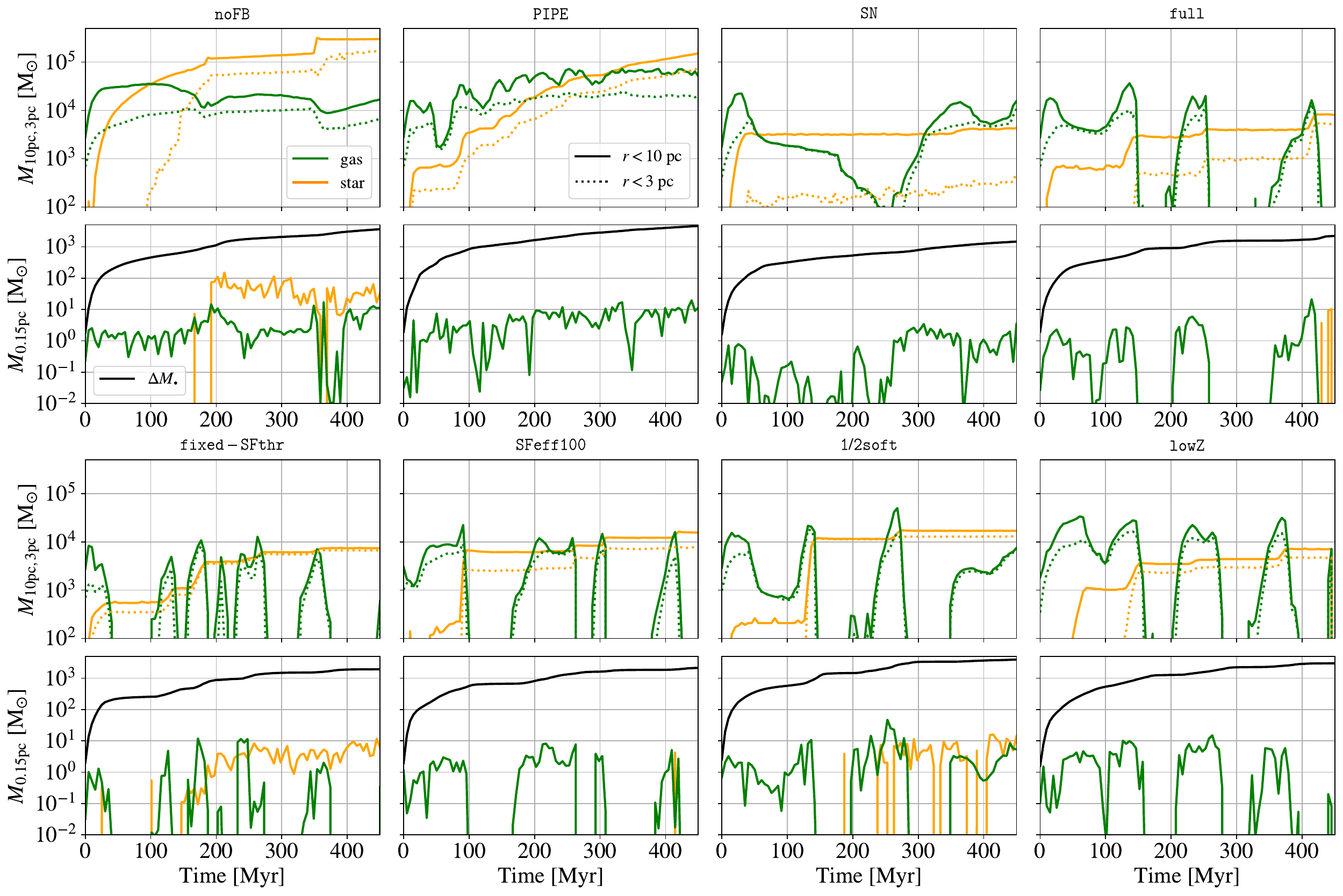}
    \caption{Time evolution of the enclosed gas mass ({\it green}) and newly formed stellar mass ({\it orange}) within $10$~pc ({\it solid} lines), 3~pc ({\it dotted} lines), and 0.15~pc. The {\it black} lines in the {\it second} and {\it bottom} panels show the mass accreted by the BH. The evolution of stellar and gas masses within the CND shows a clear dependence on the choice of stellar feedback, whereas different star formation models yield similar results due to self-regulation by feedback. In some runs, stars formed at larger radii migrate inward to the central $0.15$~pc, while in the {\tt fixed-SFthr} run, stars form directly within $0.15$~pc owing to the `relaxed' star formation criterion. See Section~\ref{sec:star-gas-mass} for details.}
    \label{fig:gas-star-evol}
\end{figure*}

\subsubsection{Impact of early stellar feedback on the CND: increased susceptibility to SNe}
\label{sec:SN-full}

To demonstrate why the CND exists only episodically in simulations that include the full feedback physics, we compare the CND surface density and density-weighted projections in the {\tt SN} and {\tt full} runs. Fig.~\ref{fig:SN-full} shows instances with comparable star formation episodes in the two simulations, where the distribution of CND mass, SFR, and the number of SNe are very similar. As shown in Fig.~\ref{fig:CNDcycle}, these episodes correspond to $\sim10$--$50$~Myr in the {\tt SN} run and $\sim110$--$150$~Myr in the {\tt full} run. For each case, we take the snapshot 8~Myr before the first SN event in the episode and follow the subsequent evolution in intervals of 8~Myr. The {\it stars} and {\it circles} indicate the locations of feedback-producing stars: in the {\tt SN} run, those that end as SNe, and in the {\tt full} run, stars producing both EUV/FUV photons and SNe. The {\it marker} size scales with stellar mass, while the colour denotes stellar age.

As shown in the first columns of the respective runs, which is 8~Myr before the onset of SNe, the stellar distribution in both simulations lies mainly near the CND edge, with only minor differences in the number of stars. However, the gas density and temperature distributions differ markedly. In the {\tt SN} run, the CND edge remains relatively dense and cold, with $T<100$~K. In contrast, in the {\tt full} run, massive stars immediately emit EUV photons, heating the CND. As a result, the CND edge where these stars are located is hotter and more diffuse, while gas is compressed inward to $\sim1$--$3$~pc from the centre, thereby enhancing BH accretion. This radiative feedback makes the CND more susceptible to subsequent SNe. Despite the CND being slightly less massive in the {\tt SN} run, it survives throughout the full simulation time of 450~Myr. By contrast, in the {\tt full} run, once the first SNe explode, the entire CND is disrupted and dispersed, leaving no disc until the next inflow into the nuclear region occurs. Thus, with early stellar feedback, the CND undergoes episodic cycles of destruction and reformation. \ej{It will be important to quantify the CND resilience and the likely cyclic nature in simulation models that include BH feedback as well.}

\subsubsection{Gas and stellar mass evolution in the CND}
\label{sec:star-gas-mass}

In this section, we quantitatively compare how the stellar and gas masses within the CND evolve at different radii across all models. The {\it first} and {\it third rows} of Fig.~\ref{fig:gas-star-evol} show the time evolution of the enclosed masses of gas and stellar masses within 10~pc and 3~pc spheres centred on the BH (see also Table\ref{tab:runs}). The {\it second} and {\it bottom rows} Fig.~\ref{fig:gas-star-evol} shows the accreted BH mass as well as the enclosed gas and stellar mass within $0.15$~pc, which is comparable to the typical self-gravity radius of the accretion disc.

Looking first at the {\it top} and {\it third} rows of Fig.~\ref{fig:gas-star-evol}, in the absence of SN feedback (i.e., {\tt noFB} and {\tt PIPE} runs), the CND remains persistently Toomre-unstable and forms stars almost continuously. As a result, by 450~Myr the total mass of newly formed stars exceeds $10^5~\msun$. After $t=150$~Myr the {\tt PIPE} run maintains the largest gas mass among all runs within 10~pc, 3~pc, and 0.15~pc, showing that BH accretion proceeds most efficiently. 
By contrast, in the other six runs that include SN feedback, star formation in the CND proceeds episodically, and the total stellar mass increases only by $\sim 10^3$--$10^4\msun$ over 450~Myr, depending on the model. When comparing the runs with only SN feedback to the full-physics runs, the CND experiences more frequent star formation episodes (see Fig.~\ref{fig:CNDcycle}), leading to a slightly increased stellar mass.

At smaller radii ($r<0.15$~pc), the mass budget is dominated by inflow toward the BH in all simulations, with the BH accreting more than $10^3~M_\odot$ over 450~Myr, regardless of the adopted model (a detailed comparison is discussed later). The comparison with Fig.~\ref{fig:3d} allows us to distinguish whether the stars within 0.15~pc formed in situ or migrated inward from larger radii. In the {\tt noFB} run, a cluster formed at larger radii merges into the nucleus at $\sim180$~Myr, depositing $\sim100~\msun$ of stellar mass into the inner CND. In the {\tt full} run at $t\sim420$~Myr and the {\tt 1/2soft} run for $t>200$~Myr, stars formed in the CND inspiral toward the centre 0.15~pc. In contrast, in the {\tt fixed-SFthr} run, stars form directly within $r<0.1$~pc, since, as shown in Fig.~\ref{fig:3d}, the more relaxed star formation criterion allows star formation to proceed even in the innermost regions.

Interestingly, among the five runs that include full physics, although detailed temporal evolution differs slightly, the overall trends in gas and stellar mass evolution are qualitatively and quantitatively similar. While the choice of stellar feedback strongly affects the circumnuclear evolution, the choice of star formation model (i.e., {\tt fixed-SFthr} or {\tt SFeff100}) has little impact on the total stellar mass, owing to the self-regulating nature of the feedback.

\subsection{Black hole accretion}
\label{sec:BHgrowth}
 \begin{figure*}
 \centering
\includegraphics[width=0.7\textwidth]{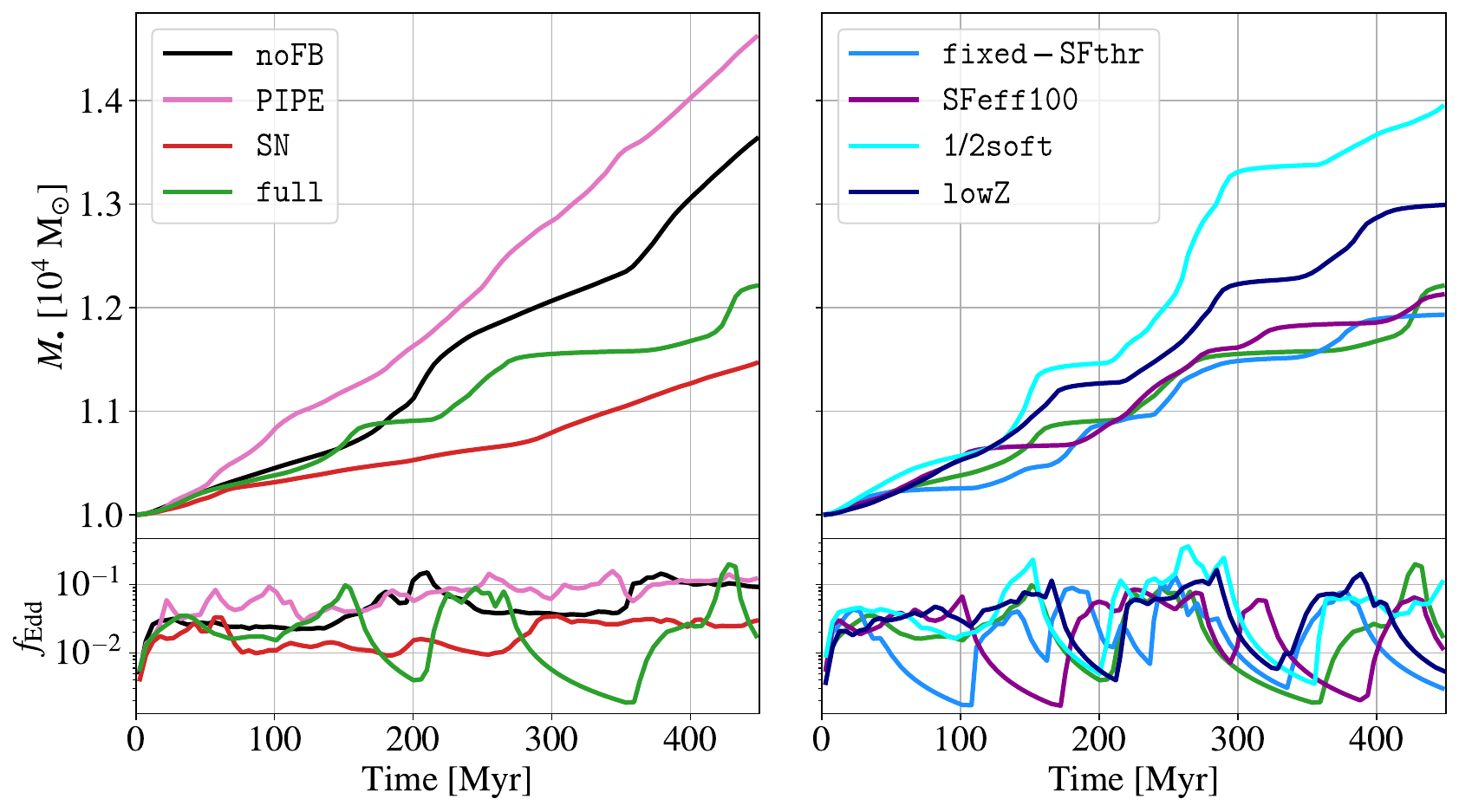}
    \caption{Evolution of the BH mass ({\it top}) and Eddington fraction ({\it bottom}) in the simulations. Stellar feedback primarily determines both the rate and mode of BH growth, i.e. whether it is episodic or continuous. However, when the feedback configuration is fixed, feedback inherently regulates the CND evolution, so the choice of star formation model has little effect. The {\tt 1/2soft} and {\tt lowZ} runs form denser and more massive CNDs, leading to faster BH growth. See Section~\ref{sec:BHgrowth} for details.
}
    \label{fig:BHgrowth}
\end{figure*}

Fig.~\ref{fig:BHgrowth} shows the time evolution of BH mass and Eddington ratio across all simulations. 
We first examine the impact of different star formation models on BH growth. Interestingly, in the simulation that includes only early stellar feedback, the BH accretes more rapidly than in the {\tt noFB} run, with a qualitatively similar conclusion reported by \citet{Petersson+2025}. This enhanced accretion can be attributed to two main factors: first, as shown earlier, the fragmentation in the circumnuclear region is suppressed, leading to the formation of a denser and more massive CND that allows for more efficient BH accretion (see Figs.~\ref{fig:3d} and \ref{fig:CNDcycle}), and second, the early stellar feedback induces additional mass inflows toward the BH.
This is clearly reflected in the evolution of the Eddington ratio: in the {\tt noFB} run, except for two rapid growth episodes around 180~Myr and 350~Myr, associated with the merger of massive clumps, the BH accretes at an almost constant rate. In contrast, the BH in the {\tt PIPE} run accretes in a more bursty pattern with shorter timescales (on the order of 10~Myr). These frequent, short-lived accretion episodes are driven by photoionisation feedback from OB stars formed within the CND, which increases the gas pressure and induces rapid inward compression of gas toward the BH.

On the other hand, in simulations that include SN feedback ({\tt SN} and {\tt full} runs), BH growth is somewhat suppressed. As discussed earlier, in the {\tt SN} run, the CND does not fully evaporate over the 450 Myr, and continuously supplies gas to the $\alpha$-disc, keeping Eddington ratio persistently above 0.01. In contrast, the accretion behaviour in the {\tt full} run differs markedly from that in the other runs. As shown in Fig.~\ref{fig:SN-full}, massive stars formed in the circumnuclear region ionise and heat the surrounding gas via radiation almost immediately after their formation. As a result, the CND becomes more susceptible to the subsequent SN feedback, ultimately leading to its eventual evaporation. Consequently, gas inflow toward the accretion disc is entirely quenched until a fresh gas supply is delivered from the larger-scale galactic disc.

Comparing the simulations in the {\it left-hand} panel of Fig.~\ref{fig:BHgrowth}, which adopt different star formation and modelling setups, we find that the BH evolution in all cases exhibits {\it episodic} growth with a period of roughly 100~Myr, similar to the {\tt full} run. This behaviour arises due to the CND being accreted and subsequently evaporated by in-situ stellar feedback. Among these, the {\tt lowZ} and {\tt 1/2soft} runs show larger BH accretion. In the {\tt lowZ} case, the low metallicity leads to less cooling, suppressing star formation and allowing the CND mass to remain higher than in the other setups. In the {\tt 1/2soft} case, the CND density remains higher than in the other simulations. As a result, in the {\tt full} run, BH accretion proceeds in short, bursty episodes following the CND evolution, and the Eddington ratio fluctuates between 0.002 and 0.2. Although the inflow toward the accretion disc is quenched for several hundred Myr, the BH ultimately grows more efficiently in the {\tt full} run than in the {\tt SN} run.

Finally, focusing on the {\it right-hand} panel of Fig.~\ref{fig:BHgrowth}, it is worth emphasising that the {\tt full}, {\tt SFeff100} and {\tt fixed-SFthr} models lead to a very similar BH growth. This is very encouraging and indicates that, regardless of the detailed star formation scheme choices, BH growth is robust. Another interesting point is that the {\tt 1/2soft} and {\tt lowZ} runs show BH growth that is nearly as efficient as in the {\tt noFB} and {\tt PIPE} runs. As discussed earlier, this is because the CND remains the most massive and dense in these two runs. This finding may also be linked to recent {\sc JWST} observations reporting compact, dense nuclear gas structures and vigorous BH growth in low-metallicity systems at high redshift \citep[e.g.,][]{Ubler+2023, Maiolino+2024}. This suggests that conditions similar to those in the {\tt lowZ} and {\tt 1/2soft} runs could naturally arise in early galaxies, promoting rapid BH growth even in the absence of strong feedback regulation.

Regarding the evolution of angular momentum in the BH–disc system, the $\alpha$-disc dominates the total angular momentum budget. The relative angle between the BH and disc remains nearly aligned (below $10^{\circ}$) throughout $450~\mathrm{Myr}$, leading to coherent accretion and the BH spin parameters increase from $a_\bullet = 0.7$ to above $0.8$ in all runs (see Appendix~\ref{sec:appendix-BHother}).

\section{Discussion}
\label{sec4:discussion}

In this study, we built upon the {\sc MandelZoom} I framework to investigate how the CND, its embedded accretion disc, and the central BH respond to variations in star formation and stellar feedback prescriptions. We found that feedback strongly reshapes the CND and its multiphase ISM, thereby regulating disc fragmentation and subsequent star formation. These processes, in turn, directly impact the mass growth and spin evolution of the central BH, demonstrating the crucial role of CND-scale physics in connecting galactic inflows to BH accretion.

The studies most closely related to ours are those of \cite{Partmann+2025} and \cite{Petersson+2025}. Our results are in broad agreement with \cite{Partmann+2025}, who model dwarf galaxies with an NSC, but without additional refinement around the BH, where, similarly to our work, they find that the presence of an NSC promotes gas accretion onto IMBHs (see \citetalias{Shin+2025} for a more detailed comparison). \cite{Petersson+2025} modelled IMBH accretion in isolated dwarf galaxies with a subgrid accretion disc model, but without incorporating an NSC. In their simulations, BH accretion occurred intermittently even in models including only SN feedback, whereas in our simulations, the CND was able to withstand SN feedback and survive, leading to continuous BH accretion over 450~Myr. We attribute these different outcomes to the absence of an NSC in their models and the adoption of a different stellar feedback injection scheme, which by construction preferentially couples feedback into the dense surrounding gas, rather than isotropically \citep[see e.g.,][]{Hopkins+2018, Smith+2018}. Hence, in \cite{Petersson+2025}, the gas near the BH may not be sufficiently dense and is thus easily disrupted by SN explosions, while in our case, despite SN events occurring tens of times more frequently within the central $50$~pc region, the CND remained structurally intact.

Interestingly, even in the absence of an NSC and thereby a lack of a gas reservoir such as a CND in \cite{Petersson+2025}, they still reported comparable or even faster BH growth, particularly in runs without feedback or with only radiation feedback. This discrepancy mainly arises from the definition of the radius at which the accretion rate is measured. The inflow rate scales with radius from the BH (see inflow rate profile in Fig.~\ref{fig:appendix-gas-profs}), such that measuring accretion farther out systematically overestimates the circularised gas inflow onto the accretion disc \citep[see also,][]{Guo+2023}. This is an issue that persists in all simulation models that cannot fully resolve the actual BH accretion disc, highlighting the critical importance of spatial resolution reached in the galactic innermost regions.

Our results extend previous observational and theoretical studies of CNDs. Whereas earlier works on CND evolution often adopted idealised models, our simulations start from a galactic-scale ISM including an NSC and follow, with non-equilibrium cooling, how gas interacts with the NSC to form a CND in a self-consistent manner. We further demonstrate how various forms of feedback, including SNe and radiation, cumulatively influence CND–BH feeding. Consistent with the observational inference by \citet{Liu+2012} and \citet{Hsieh+2017} that inflowing gas streams replenish the CND, our simulations trace the entire cycle from galactic-scale inflows to nuclear disc accretion, thereby establishing a continuous link between galaxy-scale gas supply and BH growth.

As reported by \citet{Izumi+2023}, our simulations show that the CND becomes Toomre-unstable at $r > 1$~pc, leading to star formation (see also Fig.~9 in \citetalias{Shin+2025}). They also found that roughly 3\% of the inflowing gas into the CND is further accreted onto the BH, consistent with our results: the inflow rate toward the CND is $\lesssim10^{-3}\,\msun\,{\rm yr^{-1}}$, while that toward the accretion disc is $\sim10^{-5}\,\msun\,{\rm yr^{-1}}$. The longevity of the CND has long been debated: some studies argue that it is a short-lived structure, dispersed within only a few Myr by SNe and stellar winds \citep[e.g.,][]{Requena-Torres+2012, Lau+2013}, whereas others suggest that external gas inflows from the outer disc may prolong its survival \citep[e.g.,][]{Liu+2012, Hsieh+2017}. Our simulations, starting from a galactic-scale ISM interacting with the NSC and including non-equilibrium cooling, allow us to address this question in a self-consistent manner. We find that in full-physics runs, the CND can persist from as short as 10~Myr up to more than 100~Myr, and that its life cycle is tightly linked to the in-situ SFR, which in turn controls the OB SFR and the frequency of SNe.

\ej{Over the course of the 450 Myr simulation, the NSC mass grows only marginally through intermittent in-situ star formation, by $\lesssim 3\%$ (approximately $7\times10^3$--$1.6\times10^4\, \mathrm{M}_\odot$), while the BH mass increases by $20$--$30\%$ ($\sim (2$--$3)\times10^3\, \mathrm{M}_\odot$). This differential growth leads to a modest upward shift in the $M_{\bullet}-M_{\rm NSC}$ plane, yet the evolutionary track remains well within the intrinsic scatter of the observed relation \citep[e.g.,][]{Nguyen+2018}. 
In terms of NSC structural evolution, the effective radius of the NSC shows no significant oscillations or systematic trends in most runs over 450 Myr. 
}

In the remaining paragraphs, we highlight several limitations of our model and outline possible directions for future improvements (see also \citetalias{Shin+2025}). The most important limitation is that, for the sake of simplicity, we focused exclusively on the impact of stellar feedback on BH accretion, while AGN feedback was not included. 
\ej{Although AGN feedback is not explicitly modelled in this study, the simulated accretion rates allow a rough estimate of the expected AGN luminosities assuming a spin-dependent radiative efficiency $\epsilon_{\rm r}$. For the fiducial stellar-feedback runs, the peak accretion rates correspond to bolometric luminosities in the range $L_{\rm bol} \sim 4.7\times10^{40} - 5.6\times10^{41}\,{\rm erg\,s^{-1}}$, with peak Eddington ratios of $f_{\rm Edd} \sim 0.08-0.3$. 
Recent radio observations of IMBH candidates in dwarf galaxies with comparable Eddington ratios ($\sim 0.04-0.32$) and luminosities have demonstrated detections of faint AGN \citep[see e.g.,][]{Yang+2022,Yuan+2025,Flores+2025}, revealing compact emission, pc-scale jets, variability, and in some cases powerful outflows.
Although radio detection remains challenging with current facilities, these results suggest that the IMBH in our simulations could be observable as a low-luminosity AGN under favourable conditions (e.g. jet activity or limited obscuration). Upcoming facilities such as SKAO are expected to reach substantially improved sensitivities and should enable robust detections of such faint sources in the near future.}

\ej{However, AGN feedback, such as jet, wind and radiation feedback, can strongly affect the CND and the ISM, and consequently regulate the inflow rate of gas into the galactic centre \citep[see e.g.,][]{Bourne+2014, Bourne+2015, Mukherjee+2018, Torrey+2020, Talbot+2022, Mercedes-Feliz+2023}. It remains to be understood if AGN feedback is able to significantly affect the local ISM and CND properties, and hence the mass supply onto the central BH, and in our future work, we will tackle this crucial issue.} 

Our accretion disc model follows a radiatively efficient $\alpha$-disc prescription, appropriate for geometrically thin and optically thick flows. However, when the CND is disrupted or dispersed and the accretion rate drops (e.g. with $f_{\rm Edd} \lesssim 10^{-2}$), the flow may transition toward a radiatively inefficient, geometrically thick state \citep[e.g.,][]{Yuan+Narayan2014, Koudmani+2024}. This transition not only affects the BH mass growth, but also the spin magnitude and direction evolution, which is especially important to model accurately once AGN jet feedback is considered. A more detailed treatment of this physical regime will be investigated in future extensions of this study.

Also, our simulations do not include magnetic fields, which can influence star formation and the transport of angular momentum in galactic and circumnuclear gas. Magnetic fields provide additional pressure support that can affect star formation and mediate angular momentum transport \citep{Marinacci+2018, Martin-Alvarez+2020, Hopkins+2024a}. In future work, it will be particularly interesting to constrain the effect of realistic magnetic field strengths on the star formation and fragmentation in the CND as well as on the mass flux onto the $\alpha$-disc.

Our simulations do not include a cosmological context. However, at low redshift, assuming a `field' dwarf galaxy, evolving for long periods of cosmic time in isolation is not an unreasonable approximation. Major mergers are exceedingly uncommon in such low-mass systems; for example, \citet{Koudmani+2022} found only a single minor merger event for a dwarf of comparable mass at \( z = 4 \) \citep[see also][]{Martin+2021}. In contrast, as recent JWST observations have revealed, at high redshifts dwarf–dwarf interactions and inflows of dense gas are likely much more common, making it essential to model such environmental effects to fully understand BH growth and star formation in these early Universe systems \citep[e.g.,][]{Hsiao+2023}.

Recent studies, particularly in the IMBH mass regime, suggest that BHs embedded in NSCs can grow efficiently through the TDE channel \citep[see e.g.,][]{Zubovas+2019, Alexander+2017, Rizzuto+2023, Chang+2025, Rantala+2025}. In our model, however, the NSC was represented with a single-mass population experiencing softened gravitational interactions ($\epsilon_{\rm NSC} = 0.175$~pc), which prevents us from capturing the TDE channel as well as processes such as relaxation and mass segregation within the NSC. It will be particularly timely to consider IMBH growth through TDEs in full galaxy formation simulations, which, if significant, will alleviate the difficulty of feeding (low mass) IMBH within shallow or gas-poor gravitational potentials.

\section{Conclusions}
\label{sec5:conclusion}

In this work, we carried out a suite of high-resolution simulations within the {\sc MandelZoom} framework to study how stellar feedback regulates the growth of IMBHs embedded in NSCs of dwarf galaxies. Using a super-Lagrangian refinement scheme, we resolved scales down to the self-gravity radius of the $\alpha$-accretion disc ($<0.01$~pc), enabling a self-consistent treatment of gas inflows from the resolved ISM to the CND and finally onto the BH. Within our simulation suite, we systematically explored different combinations of stellar feedback channels (photoionisation, photoelectric heating, and SNe), as well as a range of star formation prescriptions, star formation efficiencies, gas gravitational softening and initial gas metallicity values. The impact of these modelling choices on the ISM, the CND, and BH growth is summarised as follows:

\begin{itemize}
    \item \textbf{ISM:}  
    The thermodynamic structure and evolution of the ISM depend strongly on the adopted feedback model. In the absence of feedback, the disc fragments severely, triggering starbursts followed by a gradual decline in SFR due to gas depletion. With radiative stellar feedback, fragmentation is significantly suppressed, leading to smoother and more sustained star formation. SN feedback alone results in the clustered SN explosions that significantly heat and disperse the ISM. When all feedback channels are included, EUV radiation mitigates excessive SN clustering, yet the ISM develops large low-density voids while maintaining a multiphase structure and self-regulated star formation.  

    \item \textbf{CND:}  
    Compact, cold CNDs form robustly in all simulations, but the CND stability and star formation activity depend strongly on stellar feedback modelling. Without feedback or with radiative stellar feedback alone, the CND remains Toomre-unstable and forms stars continuously, accumulating more than $10^5~\msun$ of stellar mass. With SN feedback only, in-situ star formation becomes intermittent; however, the CND is not disrupted by SNe, but instead survives and continues to feed the BH throughout the entire simulation time (450~Myr). When both radiative stellar and SN feedback are included, the CND becomes intermittent: `early' feedback heats and diffuses the outer disc, making it more vulnerable to subsequent SNe, which then fully disrupt the CND. This leads to recurrent $\sim$100 Myr cycles of CND destruction and reformation.

    \item \textbf{BH growth:}  
    BH growth is closely linked to the CND cycle. Without stellar feedback, BHs grow smoothly with occasional bursts from clump–CND mergers. With radiative stellar feedback only, BH accretion becomes more bursty due to feedback-driven gas compression within the CND. With SN feedback alone, BH growth is reduced but not fully quenched since the CND survives. When all stellar feedback channels are included, the CND is disrupted entirely, and BH fueling is suppressed until fresh gas arrives from galactic scales. BH accretion then proceeds in episodic bursts with a characteristic timescale of $\sim100$~Myr.  

    \item \textbf{Impact of additional parameters:}  
    Other physical and numerical parameters also shape nuclear gas dynamics and BH growth. Unsurprisingly, a smaller gas gravitational softening length leads to gas fragmentation on smaller scales and to higher densities, causing occasionally higher BH accretion rate peaks. High star formation efficiency, or more efficient star formation prescription in the super-Lagrangian refinement region, leads to somewhat different SFRs in the CND, but qualitatively results remain unchanged, with the CND experiencing cycles of destruction and reformation leading to the episodic BH growth. Reassuringly, the overall SFR and the final BH mass are very robust to these modelling changes. Initially, lower gas metallicity suppresses radiative cooling, which reduces star formation, allowing larger gas reservoirs to accumulate in the nucleus and thereby promote higher BH accretion, with interesting implications for modelling high redshift dwarfs.  
\end{itemize}

Taken together, our results highlight that different stellar feedback channels strongly influence the ISM structure, the inflow cycle, and ultimately the life cycle of the CND, all of which play a crucial role in regulating BH feeding. This emphasises the need to accurately model and simultaneously account for the complex interplay of all key stellar feedback processes within a comprehensive framework. 

Rather than imposing an idealised CND configuration, we follow its formation and evolution self-consistently from the galactic ISM, enabling a clear and quantitative characterisation of the CND life cycle, which remains robustly constrained even when we significantly vary our star formation prescriptions. Our modelling choices likely bracket the range of plausible star formation in the CND, allowing us to constrain the amount of gas available for IMBH fuelling, which hovers around $0.01$ with peaks exceeding $0.1$ of the Eddington rate, even in our quiescent dwarf system. Our results indicate that IMBH growth may be significant in the presence of NSCs in dwarf galaxies and we expect that at high redshifts even more copious gas supply may be available to fuel IMBH activity.   

With JWST observations starting to unravel the formation of some of the lowest mass SMBH ever probed, gravitational-wave detectors probing merger remnants above $100~\msun$, and with in the near future time-domain astronomy likely unravelling a treasure trove of TDE events, now is the exciting time to finally learn about the elusive IMBH population.

\section*{Acknowledgements}
\ej{We thank an anonymous referee for the
valuable suggestions that have helped to improve our paper.} We also would like to thank Antti Rantala and Thorsten Naab for their valuable discussions and insightful feedback on this manuscript. 
E.S. and D.S. acknowledge support from the Science and Technology Facilities Council (STFC) under grant ST/W000997/1. M.A.B. is supported by a UKRI Stephen Hawking Fellowship (EP/X04257X/1). S.K. acknowledges support from an 1851 Research Fellowship awarded by the Royal Commission for the Exhibition of 1851, as well as a Junior Research Fellowship at St Catharine's College, Cambridge. This work was enabled by the University of Cambridge Research Computing Service (www.csd3.cam.ac.uk), funded by Dell EMC and Intel using Tier2 funding from the Engineering and Physical Sciences Research Council (capital grant EP/P020259/1), together with DiRAC funding from the STFC (www.dirac.ac.uk). We also made use of the DiRAC@Durham facility managed by the Institute for Computational Cosmology on behalf of the STFC DiRAC HPC Facility. The equipment was funded by BEIS capital funding via STFC capital grants ST/P002293/1, ST/R002371/1 and ST/S002502/1, Durham University, and STFC operations grant ST/R000832/1. DiRAC forms part of the National e-Infrastructure.

%%%%%%%%%%%%%%%%%%%%%%%%%%%%%%%%%%%%%%%%%%%%%%%%%%
\section*{Data Availability}
The data underlying this article will be shared on reasonable request to the corresponding author.

%%%%%%%%%%%%%%%%%%%% REFERENCES %%%%%%%%%%%%%%%%%%

% The best way to enter references is to use BibTeX:

\bibliographystyle{mnras}
\bibliography{ref} % if your bibtex file is called example.bib

% Alternatively you could enter them by hand, like this:
% This method is tedious and prone to error if you have lots of references
%\begin{thebibliography}{99}
%\bibitem[\protect\citeauthoryear{Author}{2012}]{Author2012}
%Author A.~N., 2013, Journal of Improbable Astronomy, 1, 1
%\bibitem[\protect\citeauthoryear{Others}{2013}]{Others2013}
%Others S., 2012, Journal of Interesting Stuff, 17, 198
%\end{thebibliography}

%%%%%%%%%%%%%%%%%%%%%%%%%%%%%%%%%%%%%%%%%%%%%%%%%%

%%%%%%%%%%%%%%%%% APPENDICES %%%%%%%%%%%%%%%%%%%%%

\appendix

\section{Gas radial profiles}
\label{sec:appendix-prof}

\begin{figure*}
\includegraphics[width=\textwidth]{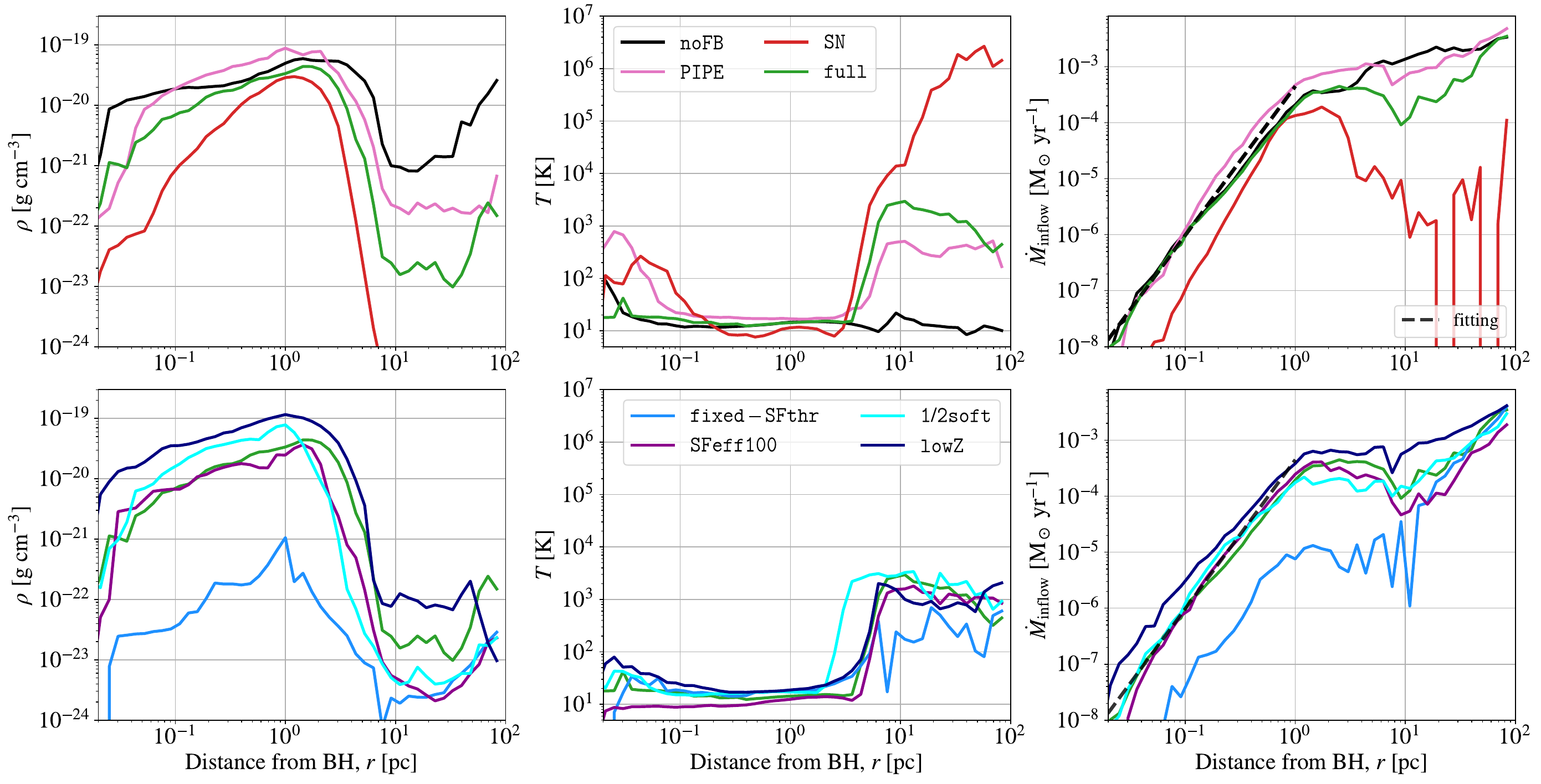}
\label{fig:appendix-gas-profs}
    \caption{Time-averaged radial profiles of gas density (density-weighted; {\it left}), temperature (density-weighted; {\it middle}), pressure (density-weighted) and mass inflow rate ({\it right}). The lines represent the median values during the first 150 Myr. The {\it thick dashed} line in the {\it right} panels shows the best-fit inflow rate profile at $r < 1$~pc for the fiducial runs excluding the {\tt fixed\mbox{-}SFthr} model. For more information, see Section~\ref{sec:appendix-prof}.}
\end{figure*}
Fig.~\ref{fig:appendix-gas-profs} presents the radial profiles of gas density, temperature, and inflow rate within 100~pc of the BH for all simulation models. Here, we show time-averaged profiles (median values computed at 1~Myr intervals) over the first 150~Myr of each run. Therefore, this figure can be regarded as representing the time-averaged results over the first 150~Myr shown in Fig.~\ref{fig:3d}. We first focus on the comparison among different stellar feedback models, and later discuss the differences from the other setups.

The {\it top-left} panel of Fig.~\ref{fig:appendix-gas-profs} displays the density-weighted gas density profiles. As seen in Figs.~\ref{fig:proj} and \ref{fig:3d}, different combinations of stellar feedback significantly impact the ISM structure in the central 10–100~pc region of the galaxy. Regardless of the feedback combination, all runs exhibit a sharp density rise at radii corresponding to the CND ($r=3$--$10$~pc), followed by a decline inside the BH’s self-gravity radius ($r\sim0.1$~pc) as the gas is depleted by BH accretion.

As expected, the {\tt noFB} run exhibits the highest gas densities throughout the galactic disc region ($r=3$--$100$~pc). In the {\tt PIPE} run, the inclusion of early stellar feedback leads to substantially reduced densities relative to {\tt noFB} across the $r=10$--$100$~pc range. Nevertheless, the central density of the CND in {\tt PIPE} remains the highest among all models. This enhancement can be understood as the result of two concurrent processes: (1) early stellar feedback suppresses star formation in the ISM, thereby permitting a greater inflow of gas toward the CND; and (2) feedback launched from the outer CND compresses the inner regions by exerting inward pressure on the gas.

The {\tt SN} run exhibits the lowest overall density, with gas densities beyond 10~pc falling below $10^{-24}\,{\rm g\,cm^{-3}}$.
The peak density of the CND is more than a factor of two lower than in the other runs, and its radius is the smallest among all models.
As shown in Fig.~\ref{fig:proj}, when only SN feedback is included, clustered SNe efficiently heat the ISM, strongly suppressing both cooling and gas inflow into the CND.
Finally, the {\tt full} run shows an intermediate density distribution between the {\tt PIPE} and {\tt SN} runs.
In particular, the gas density in the outer CND region ($r=6$–$80$~pc) is about one dex lower than in {\tt PIPE}, which can be interpreted as the result of SN feedback efficiently heating the ISM and prolonging the cooling time in that region.

The temperature distribution in the central region of the galaxy is shown in the {\it top-right} panel of Fig.~\ref{fig:appendix-gas-profs}. In the absence of feedback ({\tt noFB} run), the temperature remains very low throughout, fluctuating between 10 and 20~K. It then rises steeply near the BH, reaching $\sim$100~K around $r \sim 0.01$~pc. This temperature increase is attributed to pressure equilibrium, as the gas approaches the centre, where increasing pressure and decreasing density together result in a higher temperature.
When early stellar feedback is included ({\tt PIPE} run), the temperature in the $10$--$100$~pc region fluctuates around several hundred K. However, within the CND region ($r = 0.1$--$7$pc), the temperature stabilises at approximately 20~K. 
The {\tt SN} run shows the most extreme thermal structure as a result of supernova (SN) clustering, where the temperature in the 10--100~pc region reaches values between $10^4$ and $10^6$~K.
In contrast, the temperature within the CND drops to the lowest values among all runs, around 10~K. This is because in-situ star formation in the CND is more strongly suppressed compared to other runs (see Table~
\ref{tab:runs}), leading to minimal stellar feedback throughout most of the simulation. Additionally, with less overall gas inflow, the pressure in the central region ($r < 10$~pc) remains low, allowing the gas to maintain even lower temperatures than in the {\tt noFB} run.
\stepcounter{section}
\renewcommand{\thefigure}{\Alph{section}\arabic{figure}}
\setcounter{figure}{0}

\begin{figure*}
\includegraphics[width=\textwidth]{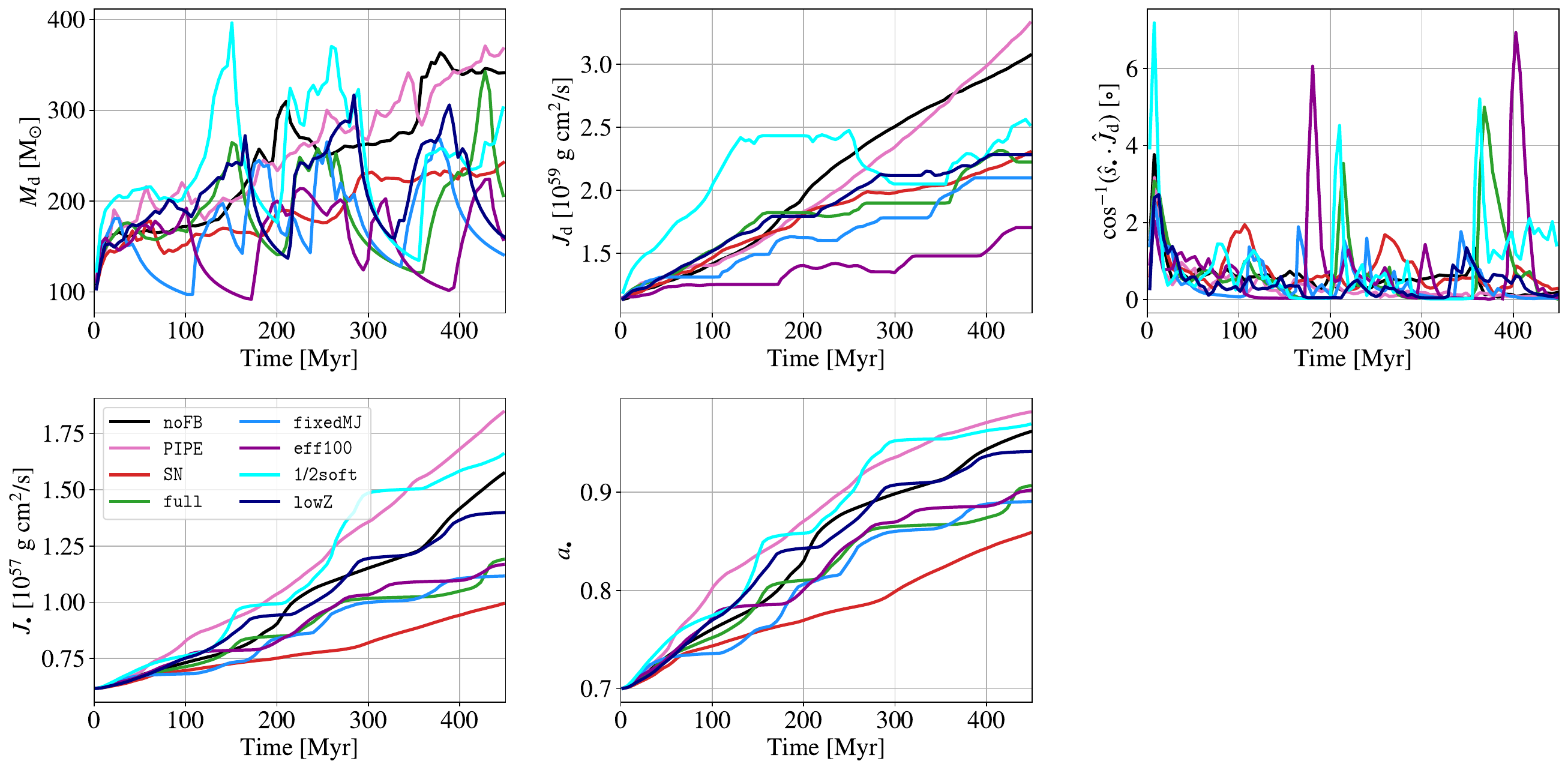}
\caption{Time evolution of the disc mass ($M_{\rm d}$; {\it top left});
the magnitude of the disc angular momentum ($J_{\rm d}$; {\it top middle});
the relative angle between the BH spin and the accretion disc ({\it top right});
the magnitude of the BH angular momentum ($J_{\bullet}$; {\it bottom left});
and the BH spin parameter ($a_{\bullet}$; {\it bottom middle}). 
For more information, see Section~\ref{sec:appendix-BHother}.}
\label{fig:appendix-BHother}
\end{figure*}

The {\it right} panels of Fig.~\ref{fig:appendix-gas-profs} shows the mass inflow rate profiles. To compute the mass inflow rate, we measure the gas mass flux across concentric spherical shells, considering only cells with negative radial velocities. Except for the {\tt SN} run, the simulations show broadly similar inflow profiles, with only minor differences in the outer few parsecs. The inflow rate increases roughly in proportion to radius within the central 1~pc \citep[also see][]{Guo+2023}, indicating that the choice of measurement radius can significantly affect the inferred BH accretion rate. This, in turn, suggests that the BH accretion rate is inevitably resolution-dependent. In this panel, we show the fitting function at $r<1$~pc for the fiducial runs (excluding the {\tt fixed\mbox{-}SFthr} model), given by \(\log_{10}(\dot{M}_{\rm inflow}/{\rm M_\odot~yr^{-1}}) = 2.66\,\log_{10}(r/{\rm pc}) - 3.35\). The inflow rate starts at approximately $10^{-8}\msun~{\rm yr^{-1}}$ near the BH (around 0.02~pc), rises to about $10^{-6}\msun~{\rm yr^{-1}}$ at 0.1~pc, around the $r_{\rm SG}$ region, and then gradually increases beyond $\sim3$~pc with mild fluctuations. As expected, the {\tt SN} run shows the most inefficient inflow due to its low ISM density. The {\tt full} run lies between the {\tt PIPE} and {\tt SN} runs, reaching inflow rates exceeding $10^{-4}\msun~{\rm yr^{-1}}$ at 10~pc.

Finally, we examine the impact of different star-formation conditions and hydrodynamic settings while keeping the same feedback configuration. The most prominent deviation is found in the {\tt fixed\text{-}SFthr} run. As shown in Figs.~\ref{fig:3d} and \ref{fig:CNDcycle}, unlike other models, this run experiences a complete disruption of the CND during the first star formation episode. The system remains without a CND for more than 50~Myr until the next inflow arrives from the galactic disc. Consequently, when averaged over time, this run shows significantly lower gas density and inflow rate compared to the other simulations. Nevertheless, as seen in Figs.~\ref{fig:3d} and \ref{fig:CNDcycle}, the CND density and SFR, when present, are comparable to those in the other models. Apart from this case, the other simulations show overall similar density, temperature, and inflow rate profiles to the {\tt full} run. However, in the {\tt lowZ} run, the lower global SFR leads to a larger galactic inflow, allowing the CND to remain the most massive among all models. In the {\tt 1/2soft} run, as seen in Fig.~\ref{fig:3d}, the CND tends to be slightly smaller in size, which can be attributed to the stronger inter-particle gravity resulting from the reduced softening length.

\addtocounter{section}{-1}
\section{accretion disc angular momentum and black hole spin evolution} \label{sec:appendix-BHother}

In this section, we describe how the accretion disc and the BH evolve over time. 
Figure~\ref{fig:appendix-BHother} presents the time evolution of the disc mass and angular momentum, the BH angular momentum and spin parameter, and the relative angle between the angular momenta of the two components. 

The evolution of the disc mass is governed by the difference between the gas inflow rate from the galactic ISM and the BH accretion rate (see Fig.~\ref{fig:CNDcycle}). Accordingly, in the full-physics runs where stellar feedback disrupts the CND, the disc mass declines rapidly, following the BH accretion rate during periods without gas inflow. In all models, the disc mass remains within the range of approximately $100-400~\msun$.

The magnitude of the disc angular momentum ($J_{\rm d}$) generally increases over time across all simulations. 
As with the mass, $J_{\rm d}$ is regulated by the balance between the angular momentum of the inflowing gas and that transported inward through accretion onto the BH.
Note that $J_{\rm d}$ typically exceeds the BH angular momentum ($J_{\bullet}$) by a factor of $100$--$300$, accounting for the majority of the total angular momentum of the BH+$\alpha$-disc system, which indicates that the torque arising from misalignment between the accretion disc and BH spin is minor.
In addition, the angular momentum transferred from the accretion disc to the BH constitutes less than 1\% of $J_{\rm d}$ (see the {\it bottom-left} panel of Fig.~\ref{fig:appendix-BHother}).
Therefore, the time evolution of $J_{\rm d}$ is mainly governed by the angular momentum supplied from the CND rather than by the internal torque within the BH-$\alpha$-disc system.

As shown in \citetalias{Shin+2025}, in the absence of an NSC where no CND forms, the inflowing gas accretes with random orientations. However, thanks to the rapid rotation of the CND, the accreted gas remains roughly aligned, although slightly misaligned (see Fig.~10 of \citetalias{Shin+2025}), leading to a gradual increase of $J_{\rm d}$ over time. One notable feature is that $J_{\rm d}$ increases more rapidly in the runs without SN feedback (i.e., {\tt noFB} and {\tt PIPE}) compared to the others. This suggests that SN feedback perturbs the direction of the inflowing gas toward the BH, weakening the coherence of its angular momentum.

Finally, regarding the alignment between the BH and the accretion disc, the relative angle between their angular momenta remains below about $10^{\circ}$ throughout the $450~\mathrm{Myr}$ evolution. Episodes of mild misalignment ($>2^{\circ}$) are mainly driven by stochastic inflows, but the Bardeen-Petterson torque quickly re-aligns the disc within a few Myr. As a result, the strong alignment between the BH and the disc enables spin-up through coherent accretion, allowing the BH spin parameter to increase from $0.7$ to as high as $0.95$ over $450$~Myr.

%%%%%%%%%%%%%%%%%%%%%%%%%%%%%%%%%%%%%%%%%%%%%%%%%%

% Don't change these lines
\bsp	% typesetting comment
\label{lastpage}
\end{document}